\documentclass{article}

\usepackage{arxiv}

\usepackage[utf8]{inputenc} 
\usepackage[T1]{fontenc}    
\usepackage{hyperref}       
\usepackage{url}            
\usepackage{booktabs}       
\usepackage{amsfonts}       
\usepackage{nicefrac}       
\usepackage{microtype}      
\usepackage{lipsum}		
\usepackage{graphicx}
\usepackage{natbib}
\usepackage{doi}
\usepackage{url}
\usepackage{lineno,hyperref}
\usepackage{amsmath,amssymb,amsfonts}
\usepackage{graphicx}
\usepackage{textcomp}
\usepackage{xcolor}
\usepackage{array}
\usepackage{url}
\usepackage{lineno,hyperref}
\usepackage{color,framed}
\usepackage{color,xcolor}
\usepackage{threeparttable}
\usepackage{multirow}
\usepackage{longtable}
\usepackage{ulem}
\usepackage{algorithm}  
\usepackage{setspace}
\usepackage{array}
\usepackage{booktabs}
\usepackage{amsmath}
\usepackage{amssymb}
\usepackage{lineno,hyperref}
\usepackage{listings,xcolor}
\usepackage{courier}
\usepackage{lineno}
\usepackage{caption}   
\usepackage{pdflscape}
\usepackage{bbding}
\usepackage{adjustbox}
\usepackage{enumerate}
\usepackage{blkarray}
\usepackage{comment}
\usepackage{blkarray}
\usepackage[super]{nth}
\usepackage{float}
\usepackage{subfig}

\newcommand{\tabincell}[2]{\begin{tabular}{@{}#1@{}}#2\end{tabular}}
\newcommand{\reffig}[1]{Fig.\ref{#1}}
\newcommand{\refeqs}[1]{Eq.\ref{#1}}

\title{A Comparative Analysis on Volatility and Scalability Properties of Blockchain Compression Protocols}

\author{ {Gerui Zhang}\\
	Center for Data Science\\
	University of Macau\\
	Macau \\
	\texttt{mc15472@um.edu.mo} \\
	\And
	{Xiongfei Zhao}\\
	Department of Computer and Information Science\\
	University of Macau\\
	Macau \\
	\texttt{yb97480@um.edu.mo} \\
        \And
        {Yain-Whar Si}\thanks{Corresponding author}\\
	Department of Computer and Information Science\\
	University of Macau\\
	Macau \\
	\texttt{fstasp@umac.mo} \\
}

\renewcommand{\shorttitle}{\textit{arXiv} Comparison}

\hypersetup{
pdftitle={A Comparative Analysis on Volatility and Scalability Properties of Blockchain Compression Protocols},
pdfsubject={cs},
pdfauthor={Gerui Zhang, Xiongfei Zhao, Yain-Whar Si},
pdfkeywords={Blockchain, Scalability, Block Compression, Transaction Fee, Block Propagation},
}

\begin{document}
\maketitle

\begin{abstract}
Increasing popularity of trading digital assets can lead to significant delays in Blockchain networks when processing transactions. When transaction fees become miners' primary revenue, an imbalance in reward may lead to miners adopting deviant mining strategies. Scaling the block capacity is one of the potential approaches to alleviate the problem. To address this issue, this paper reviews and evaluates six state-of-the-art compression protocols for Blockchains. Specifically, we designed a Monte Carlo simulation to simulate two of the six protocols to observe their compression performance under larger block capacities. Furthermore, extensive simulation experiments were conducted to observe the mining behaviour when the block capacity is increased. Experimental results reveal an interesting trade-off between volatility and scalability. When the throughput is higher than a critical point, it worsens the volatility and threatens Blockchain security. In the experiments, we further analyzed the relationship between volatility and scalability properties with respect to the distribution of transaction values. Based on the analysis results, we proposed the recommended maximum block size for each protocol. At last, we discuss the further improvement of the compression protocols.
\end{abstract}


\section{Introduction}
\textcolor{black}{Blockchain possesses desirable properties such as decentralization, robustness, transparency, and audibility. These qualities make it a suitable platform for a wide array of applications including supply chain management, commodity trade finance, banking, insurance, energy grid, and more. Despite Blockchain's attractive features and increasing variety of application scenarios, scalability is one of the most concerned aspects of Blockchain that could hinder its future development. Recent Blockchains have limitations in processing transactions per second. For example, Bitcoin can handle up to 7 transactions per second, and Ethereum can handle up to 15 transactions per second. By comparison, a conventional centralized payment system such as VISA processes 1,700 transactions per second on average. This limitation on scalability remains a major challenge for wider applications of Blockchain technology in the future.}

\textcolor{black}{In recent years, various scalability solutions were proposed in literature such as Sharding \cite{luu2016secure}, Directed Acyclic Graph (DAG) \cite{pervez2018comparative}, and Lightning Network \cite{poon2015Bitcoin}. These solutions can be divided into two layers. The former layer includes Sharding, DAG and Bigger block \cite{garzik2015block}. The latter layer includes Payment Channels \cite{poon2015Bitcoin} and Side Chains \cite{poon2017plasma}. The first-layer schemes also include compression protocols that apply compression techniques to reduce the size of the transaction during propagation. With these solutions, block capacity can be scaled to contain 60 times more transactions than before without compromising the decentralization. Meanwhile, transaction fees are gradually increasing in proportion to miners' rewards as coin-based block incentives are reduced to a negligible amount.}

\begin{figure}[htp]
    \makeatletter
    \def\@captype{figure}
    \makeatother
    \centering
    \includegraphics[width=0.6 \textwidth]{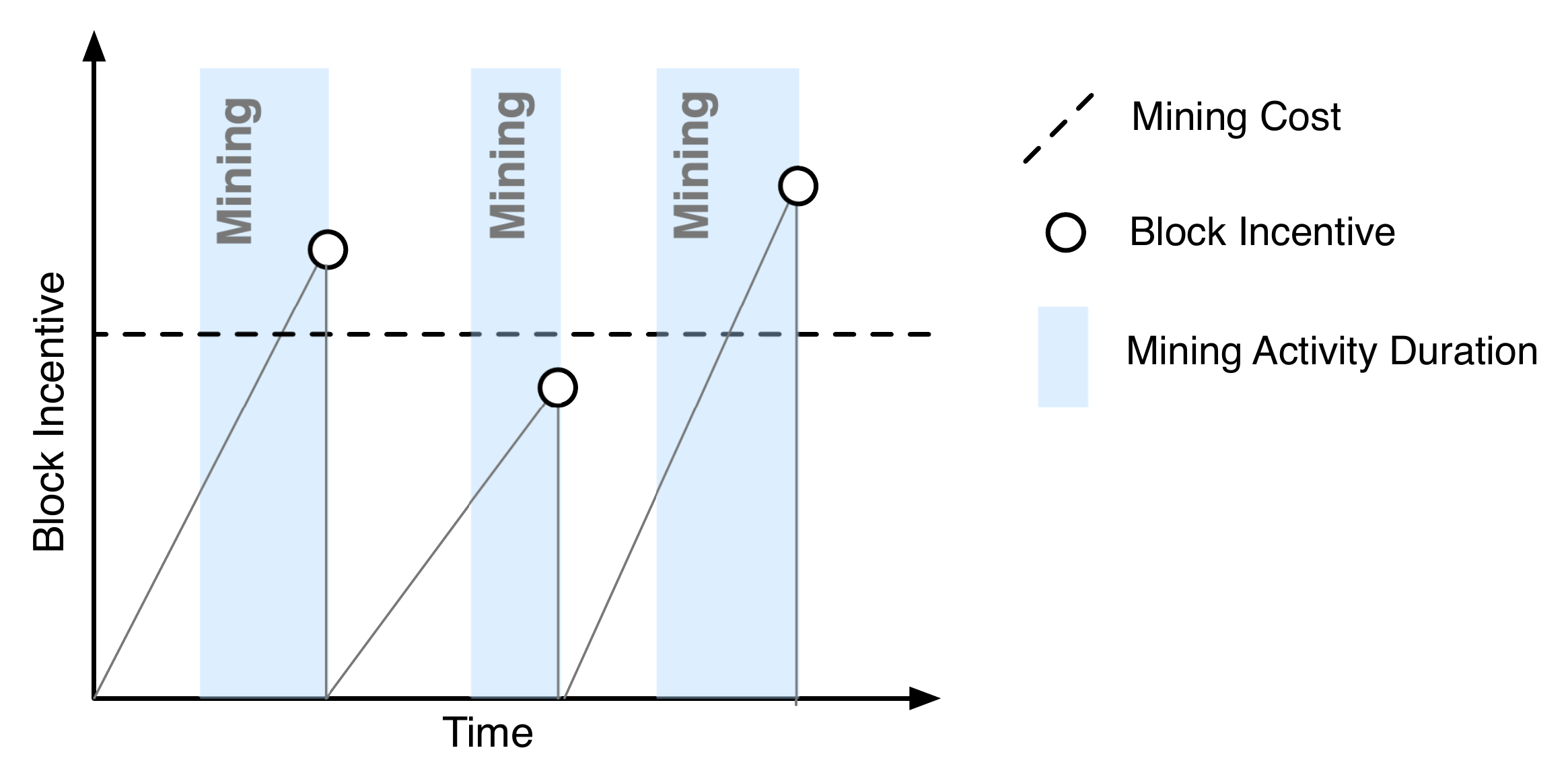}
    \caption{Illustration of Mining Gap}
    \label{MiningGap}
\end{figure}

\textcolor{black}{Carlsten et al. \cite{2978408} argued that under the transaction fees regime, deviant mining strategies such as \textsl{Selfish Mining}, \textsl{Undercutting}, \textsl{Mining Gap}, and \textsl{Pool Hopping} could hurt the mining system. In \textsl{Mining Gap}, miners need to bear the high costs of running their mining rigs and can make a profit if and only if the mining incentives exceed the total cost. With high volatility in miners' rewards, the expected reward from mining may not be able to cover the mining costs, making it unprofitable for any miner to mine. As illustrated in \reffig{MiningGap}, miners will only mine when the instantaneous expected reward exceeds the cost. \textsl{Mining Gap} can cause the effective hashing power of the Blockchain network to drop, making it easier for the malicious miners to fork.}

\textcolor{black}{The block compression protocols can increase the number of transactions that can be incorporated into the block. Due to the exponential distribution of transaction arrival time, the volatility of miners' rewards can be high, exacerbating deviant mining behaviours. Vitalik Buterin, a co-founder of Ethereum, identified the scalability trilemma which involves balancing attributes of security, decentralization, and scalability \cite{Trilemma}. According to~\cite{Trilemma}, it is difficult to achieve all three attributes simultaneously. Even though block compression protocols can improve Blockchain scalability while preserving decentralization, the solutions face serious security threats under the transaction fee regime.}


\textcolor{black}{In recent years, researchers have investigated the scalability issue from many different perspectives. However, to the best of authors' knowledge, prior research has not extensively compared the compression protocols, especially about the relationship between scaling of the Blockchain and deviant mining strategies which are caused by the increase in the volatility of fee-based miners' rewards. To this end, in this paper, we rigorously compare six high-performance compression protocols, namely Compact \cite{Compactblock}, XThin \cite{Thinblocks}, Graphene \cite{ozisik2019graphene}, XThinner \cite{XThinner}, IPFS Model\cite{zheng2018innovative}, and Dino \cite{hu2022dino}. We focus on the volatility issue caused by the scalability improvement. We conducted two sets of experiments: the first experiment evaluates the block size of Graphene and XThinner when the block capacity is increased from 1000 to 1,000,000. The second experiment simulates the mining behavior when the block capacity is increased in the same range. We then compared the performance of different compression protocols by using Transaction Per Second (TPS) as a benchmark, while maintaining the volatility of miners' rewards within the historical range. The contributions of this work can be summarized as follows:} 


\begin{enumerate}
    \item\textcolor{black}{We review six state-of-the-art compression protocols for scaling the Blockchain. In the experiments, we compare their compression rates and block capacity. }
    
    \item\textcolor{black}{For the six compression protocols, we further analyze the relationship between the scalability and the threats caused by the volatility of miners' rewards under the transaction fee regime. \textcolor{black}{We also define two terms: underscaling and overscaling to illustrate the relationship.}}
    
    \item\textcolor{black}{In this paper, we propose the recommended maximum block size for each compression protocol when the volatility of miners' rewards is maintained within the historical range. A protocol with a block size exceeding the recommended value can threaten the security of the Blockchain under the transaction fee regime. } 
    
        \item\textcolor{black}{Based on the analysis results, we further suggest the future direction for compression protocols.}
\end{enumerate}

\textcolor{black}{The rest of this paper is organized as follows. Section 2 reviews state-of-the-art scalability solutions. The background information, such as the transaction-fee regime and basic Blockchain concepts, are introduced in Section 3. Section 4 details the state-of-the-art compression protocols. The experimental results are analyzed in Section 5. This paper is concluded in Section 6.}

\section{Related Work} 

In this section, we briefly review different scalability solutions for Blockchains. In \cite{croman2016scaling}, Croman et al. proposed Maximum Throughput, Latency, Bootstrap Time, and Cost per Confirmed Transaction (CPCT) as four experimental measurements to characterize the resource costs and performance of the Bitcoin network. Croman et al. \cite{croman2016scaling} further categorized the solutions into five planes: network, consensus, storage, view, and side plane. In  \cite{hafid2020scaling}, Hafid et al. divided the scalability solutions into the first-layer (on-chain) solutions 
 and the second-layer (off-chain) solutions.

In any Blockchain, two parameters (block time and block size) are crucial for controlling the throughput. The block time governs the frequency of adding new information to the Blockchain. The block size governs the number of transactions that can be added to the Blockchain each time. Therefore, one of the simplest and most straightforward approaches to scale the Blockchain is adjusting the block size and block time. Bitcoin Cash (BCH) \cite{Bitcoincash} is a recent proposal which increases the block size to 32 MB. However, Croman et al. \cite{croman2016scaling} pointed out that block size scaling is limited. Given other parameters, such as block time unchanged, the block size should not be bigger than 4 MB.

Since the most straightforward method of increasing the block size is no longer feasible, efforts have been made in many different directions. In the network plane, a number of compression protocols such as Compact \cite{Compactblock}, XThin \cite{Thinblocks}, Graphene \cite{ozisik2019graphene}, XThinner \cite{XThinner}, IPFS Model\cite{zheng2018innovative}, Dino \cite{hu2022dino}, etc. were proposed. Compression protocols can reduce the data size in block propagation to save bandwidth and carry more transactions in a block. Recent literature on newly-proposed compression protocols only focuses on certain properties, such as compression performance. For example, in \cite{ozisik2019graphene}, Ozisik et al. compared the compressibility of Compact with Graphene based on simulation experiments. Toomim \cite{toomim_medium}, the author of XThinner, mentioned that although Graphene blocks are smaller than XThinner blocks, Graphene only has a 41\% success rate in a two-day stress test, and its recovery cost is high. Hu and Liu \cite{hu2022dino} also compared the bandwidth usage of Dino, Graphene, Compact, and XThin, particularly when certain transactions are missing. Experimental results revealed that Dino performed better than the other three competitors. 

While compression protocols can play an important role in scaling the Blockchain in the future, the research is still ongoing, and we have yet to see them widely adopted. Since compression protocols are promising for future development, we believe that compression protocols need to be considered in conjunction with the future conversion of mining incentives. Overall, the existing comparison results for compression protocol are insufficient and lacks foresight for future development.

\section{Background}
In this section, we first introduce the background of the scalability and volatility issues. We then describe four key concepts that are central to Blockchain technology. These concepts include transactions, blocks, block propagation, and miners' rewards.

Currently, Bitcoin generates a block every 10 minutes with an average block capacity of 2100. Miners participating in the Blockchain network receive users' transactions and package them into a block with a maximum size of 1 MB. By following the consensus protocols such as proof-of-work (PoW), miners compete for the right to add a new block to the Blockchain. Nevertheless, low network throughput limits the scalability of the Blockchain and undermines its stability. Additionally, long propagation time also increases the forking rate and orphan rate.

In addition to the scalability issue, the upcoming transaction-fee regime poses significant threats to the stability of the Blockchain. In the current public Blockchains, miners work for two kinds of revenue: block rewards and transaction fees. The former takes a larger proportion. However, the block reward is set to be halved whenever 2100 blocks are mined. This usually occurs approximately every four years, starting from 50 BTC in 2008. With the block reward set to reduce steadily, the transaction fee will become more important for the miners' rewards. 
	
Except for the time-varying nature of transaction fees, insufficient scaling will also result in the volatility of miners' rewards. The transaction fee in Bitcoin not only serves as a means of compensating miners but also incentivizes faster processing, especially when Bitcoin's throughput is limited. If the throughput is scaled insufficiently, the increasing block capacity can make miners rely more on transaction fees. Moreover, this insufficient scaling could increase the competition for fast processing, resulting in more volatile rewards. Carlsten et al. \cite{2978408} noted that the following strategic deviations are directly related to the stability of miners' rewards.

\begin{itemize}
    
    \item\textbf{\textsl{Selfish Mining:}} \textcolor{black}{In the transaction-fee regime, \textsl{Selfish Mining} makes slightly higher profits than it does in the block-reward regime. However, miners can decide whether to hide a new block based on their rewards in the transaction fee regime. Based on this phenomenon, Carlsten et al. \cite{2978408} proposed an improved \textsl{Selfish Mining}, which is easier to be deployed and more profitable than the original approach. Selfish miners can profit immediately when they apply \textsl{Selfish Mining}.}
    
    \item \textcolor{black}{\textbf{\textsl{Undercutting Attack:}} The attacker forks the chain and leaves more available transactions. Under the transaction fee regime, miners pursue a higher profit and are more likely to be lured by the chain that offers more rewards. The \textsl{Undercutting} makes the Blockchain vulnerable to 51\% attacks.}
    
    \item \textbf{\textsl{Mining Gap:}} \textcolor{black}{Before finding a block, miners must bear the electricity costs to mine the block. However, \textsl{Mining Gaps} can emerge without block rewards in the transaction-fee regime. The unstable rewards gained from a block may not cover miners' costs. As a result, miners' confidence in mining could be affected. Miners will only work when their expected rewards surpass their costs, which can lead to reduction in hashing power.}
    
    \item \textcolor{black} {\textbf{\textsl{Pool Hopping:}} Different mining pools have different distribution schemes. In some situations, a pool may be more successful in mining high-value blocks. Pool-hoppers often hop in and out of the same pool or to another pool to maximize their profits. Under the transaction fee regime, miners may also adopt a similar strategy to maximize their profits.} 
    
    
\end{itemize}

\subsection{Transactions}

\begin{figure}[htbp]
    \centering
    \subfloat{
        \label{Transaction}
        \includegraphics[width=0.4\textwidth]{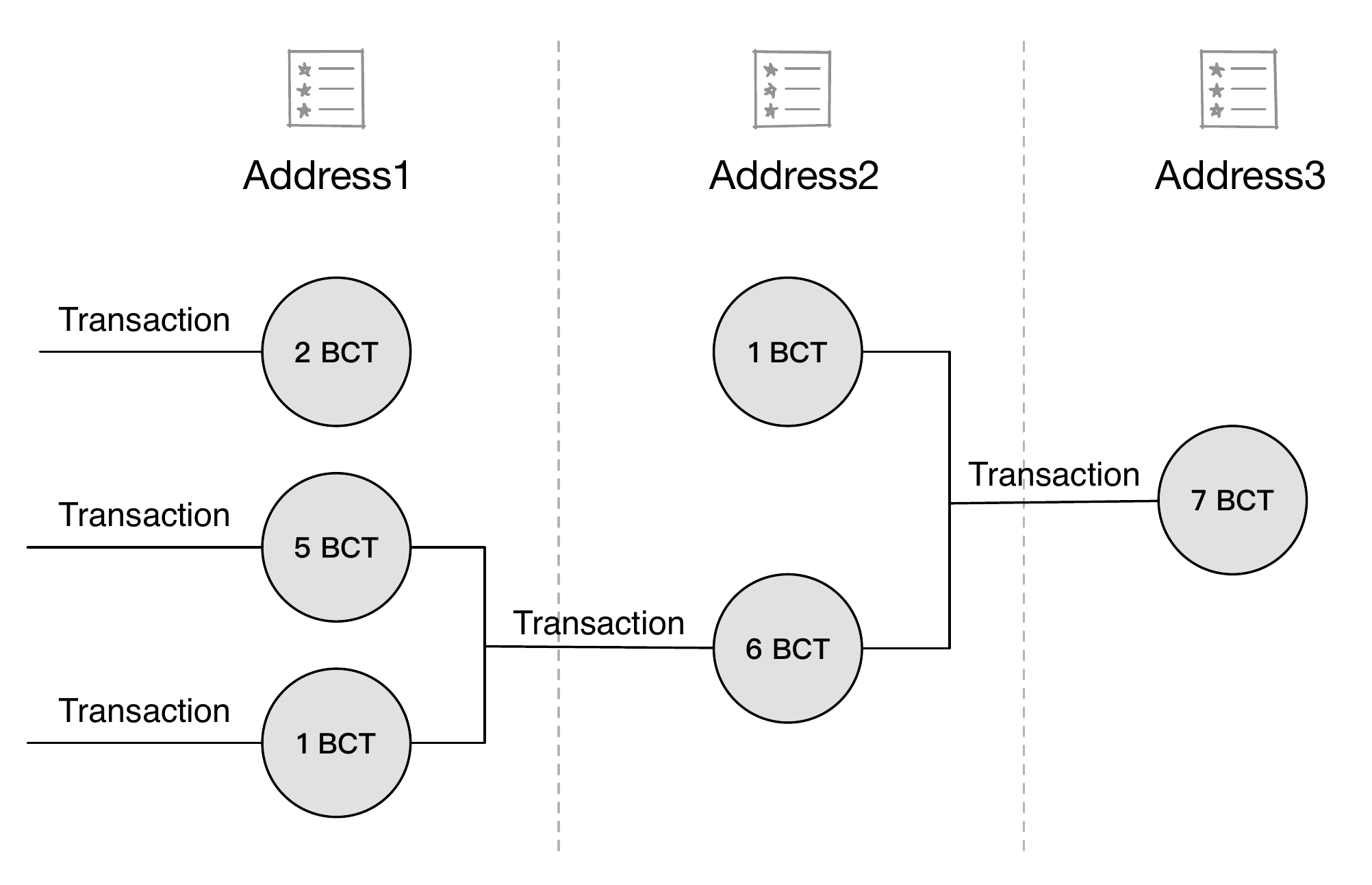}}
    \hfill
    \subfloat{
        \label{BlockSimple}
        \includegraphics[width=0.4\textwidth]{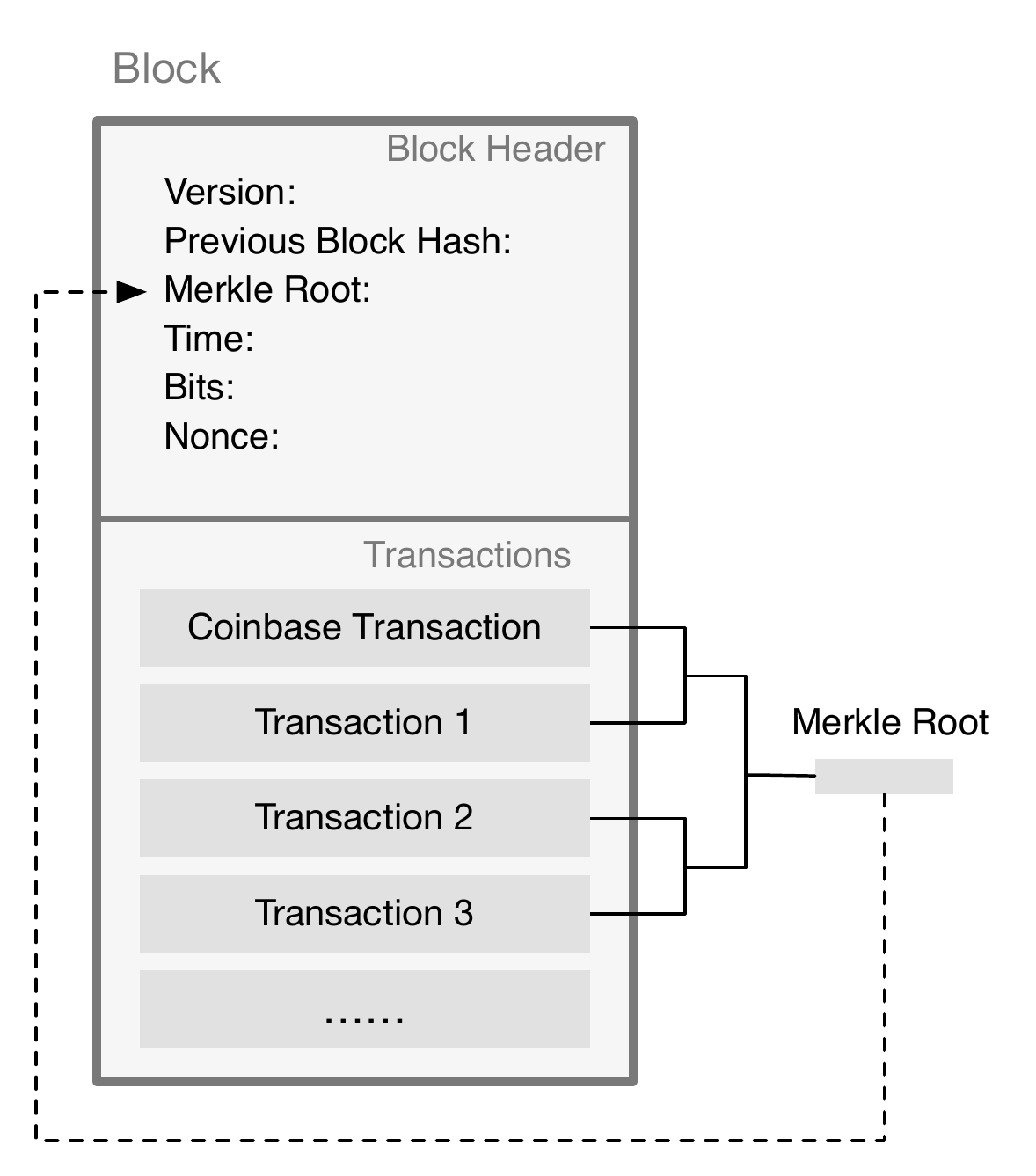}}
    \caption{(A)Illustration of the transfer of cryptocurrency (B) Structure of a block in the current Bitcoin network}
    \label{fig:block_and_transaction}
\end{figure}

Transactions are the most important part of any Blockchain network. One of the common types of Blockchain transactions is the transfer of cryptocurrency from one account to another. Fig.\ref{fig:block_and_transaction}(A) illustrates the transfer of Bitcoin cryptocurrency. In this figure, the balance of $address_1$ is the sum of the values of all unspent outputs for $address_1$. If $address_1$ wants to send $address_2$ cryptocurrencies, it sends $address_2$ a new amount based on the total balance it has received. Normally, Bitcoins are received in batches, and then new batches are created to send Bitcoin to other addresses. \textcolor{black}{The amount of cryptocurrencies sent by $address_1$ is recorded as the input value in a Bitcoin transaction, and the amount received by $address_2$ is the output value. The difference between the input value and the output value is called the transaction fee, which is earned by miners.}

\subsection{Block}

A Blockchain contains a chain of blocks, and the newer block is built on top of the previous one. Each block consists of a header and a list of transaction data arranged in a tree structure. In the block header, a hash pointer points to the previous block, and a Merkle root points to the transaction data, which allows the miners to verify that the chain sequence or transaction value has not been tampered with. Transactions are first stored in a transaction pool and then incorporated into a candidate block. The miners then attempt to add the candidate block to the Blockchain.

\reffig{fig:block_and_transaction}(B) illustrates a block's data structure in a Blockchain. The block header is the metadata at the top of the block, and the fields in the block header provide a unique summary of the entire block. The block header contains the following information, \textsl{``Version''} represents the version number of the block. \textsl{``Previous Block Hash''} is the hash summary of the last block, which is the key to connecting blocks. \textsl{``Merkle Root''} is the root of the Merkle tree after all transactions in the block were hashed together. It can also be considered a summary of all transactions in the block. \textsl{``Time''} is used to indicate when the block was mined in Unix format. \textsl{``Bits''} is a shortened version of the hash value which is the upper bound for the hash of the current block (i.e. target). \textcolor{black}{While mining a new block, miners try to adjust \textsl{``Nonce''} to make the hash of the block header lower than the target. The size of a block header is generally 80 Bytes. }

\subsection{Block Propagation}
A distributed ledger requires periodical synchronization to update the transactions and blocks. To disseminate transactions over the P2P network,  Blockchains such as Bitcoin, Bitcoin Cash, and Ethereum apply the Gossip protocol~\cite{lin1999directional} or its variation. Each node in the network maintains a local memory pool (mempool) in which new valid transactions are stored temporarily until they are confirmed and packaged into a valid block. 

Bitcoin applies two kinds of propagation protocols, namely Legacy Block Propagation protocol and Compact Block Propagation protocol which are now supported by more than 98\% of nodes in Bitcoin \cite{shahsavari2020theoretical}. The Compact Block Propagation protocol~\cite{Compactblock} transmits a Compact block which is reconstructed by the receiver with transactions in its mempool. In comparison, the whole block is transmitted over the network under the Legacy Block Propagation protocol. 

In block propagation, when a new block message is received, the recent receiver validates the block first and sends an inventory ($inv$) message to all connected nodes. Some nodes may not respond because they have received the new block. The other nodes without the block or transactions reply with a $getdata$ message to ask for the new block or transactions. After receiving the $getdata$ message, the sender sends back the required block or transactions to the receiver. In a round of Gossip communication between two connected nodes, three messages are transmitted to relay a new block. Communication between nodes varies in different compression protocols. 

\subsection{Miners' rewards}

Miners' rewards include two parts: block reward and transaction fee. Block rewards are awarded in cryptocurrency to the first miner who solves a complex mathematical problem and creates a new block of verified transactions. And to a lesser extent, transaction fees are also part of the payment to miners for mining a new block. 

Transaction fees act as an incentive for miners to incorporate transactions in their candidate block. If the total number of transactions in the mempool exceeds the number of transactions that can be incorporated into a block, transaction fees can be used as a way of prioritizing the incorporation of those transactions. When the miners fill a block with transactions, they will try to maximize the transaction fees that can be earned from the mining. Thus, transactions that offer higher fees are usually selected for inclusion in the block.

To curb inflation, Satoshi Nakamoto designed Bitcoin to be just 21 million Bitcoins in total. After the creation of every 210,000 Bitcoin blocks, roughly every four years, the amount of Bitcoin block rewards is halved. When Bitcoin started running in 2009, each block was awarded 50 Bitcoins. As of May 2021, there were already 18.7 million Bitcoins in circulation, accounting for nearly 90\% of the total planned supply. As a result, transaction fees are expected to be a major revenue for Bitcoin miners in the near future.

\section{Compression Protocols} \label{Compression Protocols}

\textcolor{black}{In this section, we review six popular block compression protocols: Compact, XThin, Graphene, XThinner, IPFS Model, and Dino. Table.\ref{ScalabilitySolutions2} lists the general descriptions and membership determinations of the six compression protocols. In the following subsections, we briefly review their mechanisms.} 
	
\textcolor{black}{Note that the compression rate of Compact, XThin, and IPFS Model are constant and provided by the original specifications of these protocols. XThin and Graphene use a probabilistic data structure called the Bloom Filter, which is detailed in the Appendices of this paper. In the experiments, we used the Monte Carlo simulation to calculate the Graphene and XThinner's block capacity to compare with others. Besides, unlike the rest of the protocols analyzed in this paper, Dino transmits block construction rules rather than compressed block content. Therefore, we will  separately discuss about Dino.}



\linespread{1.2}
\begin{table}[htp]
	\centering
	\caption{\textcolor{black}{Six compression protocols from the first layer of scalability solutions}}
	\label{ScalabilitySolutions2}
	\begin{tabular}{p{1.8cm}<{\centering} |p{2.2cm}<{\centering} |p{7cm}<{\centering}}
		\toprule
		\specialrule{0em}{3pt}{1pt} 
		\textbf{Compression protocols} & \textbf{Membership Determination} & \textbf{Description} \\
		\specialrule{0em}{3pt}{1pt} 
		\hline
		\specialrule{0em}{3pt}{1pt} 
		Compact \cite{Compactblock} & Probabilistic & A \textsl{cmpctblock} carrying 6-Byte transaction IDs is propagated.\\
		\specialrule{0em}{3pt}{1pt}
		\hline
		\specialrule{0em}{3pt}{1pt} 
		XThin \cite{Thinblocks} & Probabilistic & A \textsl{Thinblock}, which carries 8-Byte transaction IDs and missing transactions predicted using a Bloom Filter, is propagated.\\
		\specialrule{0em}{3pt}{1pt}
		\hline
		\specialrule{0em}{3pt}{1pt}
		Graphene \cite{ozisik2019graphene} & Probabilistic & \textcolor{black}{A Bloom Filter and an IBLT are propagated to the receiver, who then generates the other IBLT to reconstruct the new block.}\\
		\specialrule{0em}{3pt}{1pt}
		\hline
		\specialrule{0em}{3pt}{1pt}  
		XThinner \cite{XThinner} & Probabilistic & A \textsl{xtrblk} carrying shortened IDs around 2 Bytes is propagated.\\
		\specialrule{0em}{3pt}{1pt}  
		\hline
		\specialrule{0em}{3pt}{1pt}  
		IPFS Model\cite{zheng2018innovative} & Deterministic & A block carrying IPFS IDs which are used to download raw transaction data from IPFS is propagated.\\
		\specialrule{0em}{3pt}{1pt}  
		\hline
		\specialrule{0em}{3pt}{1pt}  
		Dino \cite{hu2022dino} & Deterministic & Nodes keep periodical communication. A Dino block which carries construction rules is propagated.\\
		\specialrule{0em}{3pt}{1pt}  
        \toprule
	\end{tabular}
\end{table}

\textcolor{black}{Theoretically, a 1MB Bitcoin block usually contains 2100 transactions. Therefore, the average size of a transaction is 500 Bytes. A coin-based transaction is also assumed to occupy 500 Bytes. We can simplify the structure of a block into four parts, namely, a block header (80 Bytes), a coin-based transaction (500 Bytes), compressed transaction data (variable), and additional block information (variable). In our analysis, \textcolor{black}{\textsl{Block Capacity} is defined as the maximum number of transactions that can be packaged in a block.} The block capacity can be calculated based on \refeqs{capacity}.}

\begin{equation}
    \textcolor{black}{Block\ Capacity = \frac{Block\ Space}{Transaction\ Size} \label{capacity}}
\end{equation}

\textsl{Block Space} refers to the space where compressed transaction data is stored, excluding the fixed space for the block header, coin-based transactions, and additional information. \textsl{Transaction Size} is the average size (in Bytes) of transactions after compression. 

\subsection{Compact Blocks} 

Compact Blocks (Compact) \cite{Compactblock} was proposed to compress the block size in Bitcoin. Compact accelerates the propagation of new blocks across the network by sending shortened transaction IDs. Upon receiving the shortened transaction IDs, receivers will compare IDs in the local mempool with the transmitted IDs to filter out the existing transactions. Only transaction data that do not exist locally are downloaded. 

\begin{figure}[H]
    \makeatletter
    \def\@captype{figure}
    \makeatother
    \centering
    \includegraphics[width=0.8 \textwidth]{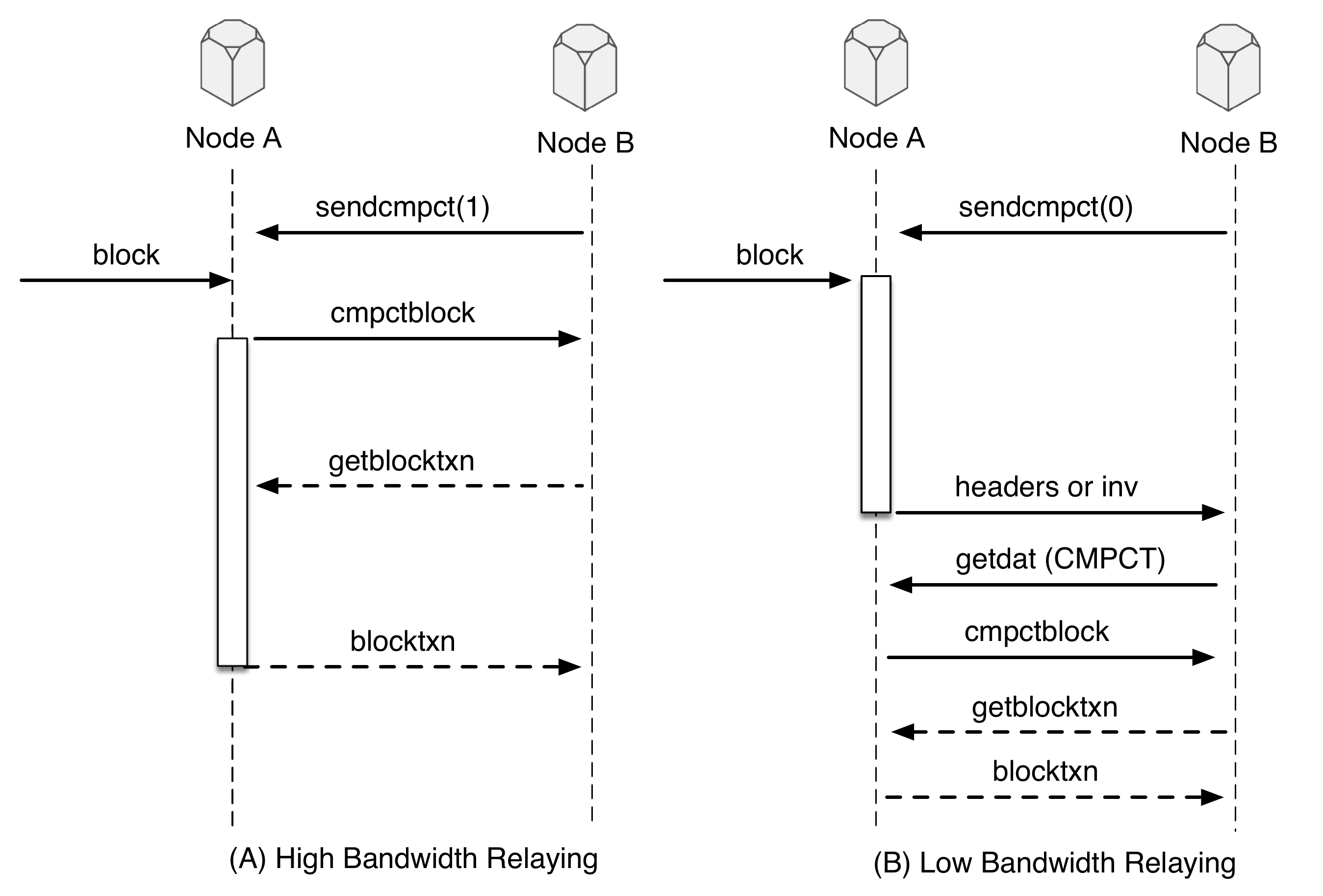}
    \caption{Compact synchronizes block information mechanism between different nodes under (A) High Bandwidth Relaying mode, and (B) Low Bandwidth Relaying mode.}
    \label{Compactblock}
\end{figure}

As shown in \reffig{Compactblock}(A), Node B uses the \textsl{sendcmpt(1)} message to inform Node A that it wants to accept the block data in Compact mode. When a new block arrives, Node A sends the block header and compressed transaction IDs to Node B. Node B compares transaction data and requests the transactions it does not have through a \textsl{getblocktxn} message. \reffig{Compactblock}(B) illustrates how Low Bandwidth Relaying is performed, which is the same as High Bandwidth Relaying. Before sending the \textsl{cmpctblock} message, Node B is informed of the new block header information by \textsl{inv} message. Node B checks whether the block already exists and then requests Node A to send the \textsl{cmpctblock} message. Low Bandwidth Relaying can further reduce data transmission and bandwidth usage in a short period.

\textcolor{black}{The \textsl{cmpctblock} sent by Node A usually includes} a \textsl{HearderAndShortID} data structure which contains a \textsl{block header} (80 Bytes), a \textsl{nonce} (8 Bytes), a \textsl{shortids length} (3 Bytes), a \textsl{prefilledtxn length} (3 Bytes), \textsl{shortids} (6 Bytes each), and \textsl{prefilledtxn}. Compact is a probabilistic approach. Therefore, the \textsl{prefilledtxn} value can vary due to hash collision. To simplify our experiments, we assume that only coin-based transaction is filled into the structure in advance. In that way, we can estimate that the Compact's block capacity is approximately 174,663.

\subsection{Xtreme Thinblocks} 

Xtreme Thinblocks (XThin)~\cite{Thinblocks} is an efficient block relay strategy. It utilizes a Bloom Filter to synchronize different miner nodes' mempool information. Under this mechanism, rather than transmitting the entire block, miners only need to transmit the transaction IDs and rebuild the full block based on the transaction data in its mempool. XThin designs two round trips to cater to all the conditions. 

\begin{figure}[H]
    \makeatletter
    \def\@captype{figure}
    \makeatother
    \centering
    \includegraphics[width=0.8 \textwidth]{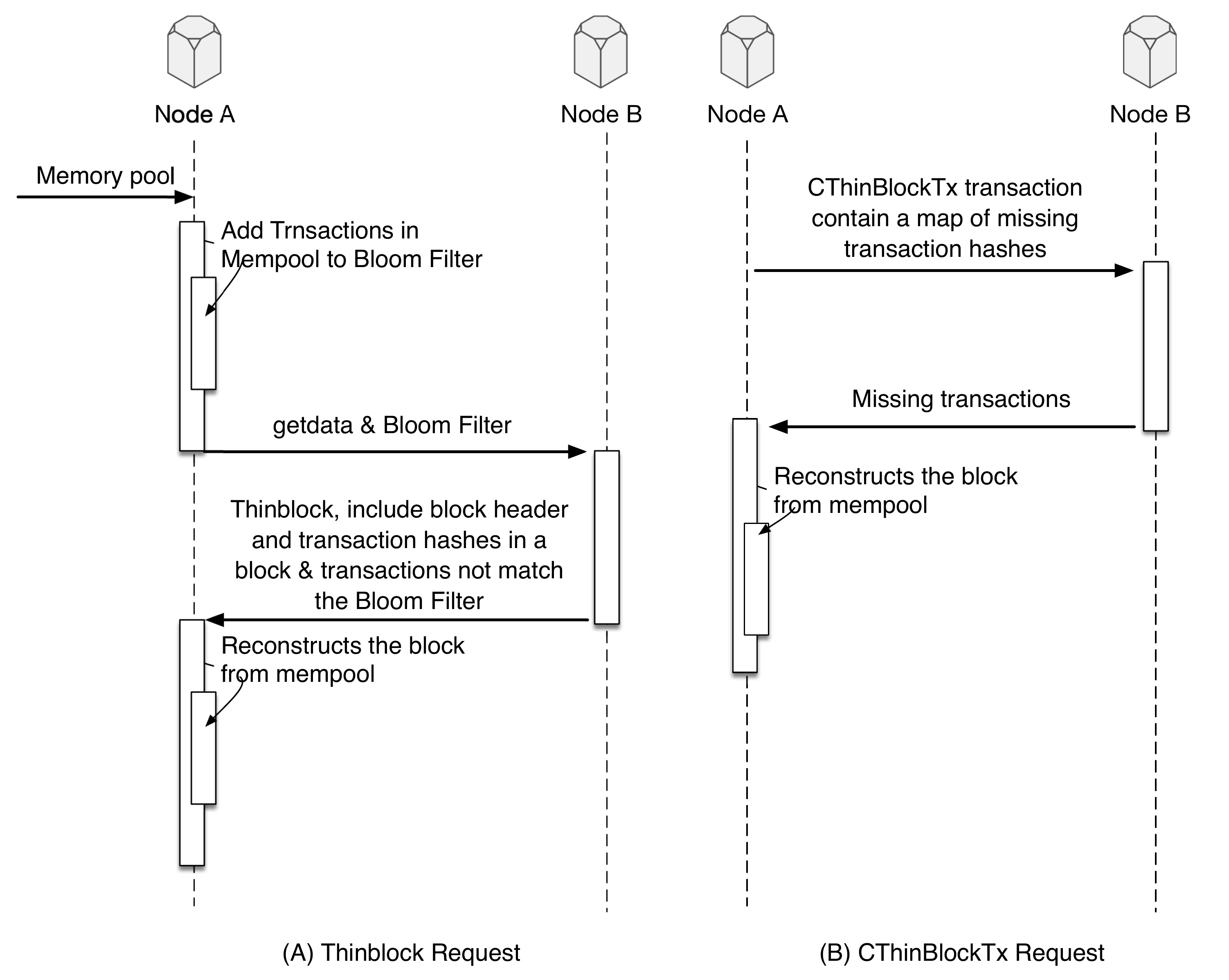}
    \caption{(A) Xtreme Thinblocks (XThin) mempool synchronization mechanism through Bloom Filter and (B) repair mechanism for missing transactions.}
    \label{Thinblocks}
\end{figure}

As shown in \reffig{Thinblocks}(A), Node A generates a Bloom Filter and encodes all transaction IDs in its mempool. When Node B receives the \textsl{getdata} request from Node A, it sends back a \textsl{Thinblock} containing only the block header, shortened transaction IDs, and transactions that are not matched by the Bloom Filter. The process shown in \reffig{Thinblocks}(B) is used to remedy the situation where Node A misses certain transactions in the block. Finally, Node A is able to reconstruct the block from its mempool.

In this paper, we assume that a \textsl{Thinblock} includes a \textsl{block header} (80 Bytes), a \textsl{prefilled transaction} (500 Bytes) and a list of 8-Byte transaction hashes. Based on this assumption, we can estimate that the block capacity is approximately equal to 130,999. 

\subsection{Graphene} 

\begin{figure}[H]
    \makeatletter
    \def\@captype{figure}
    \makeatother
    \centering
    \includegraphics[width=0.8 \textwidth]{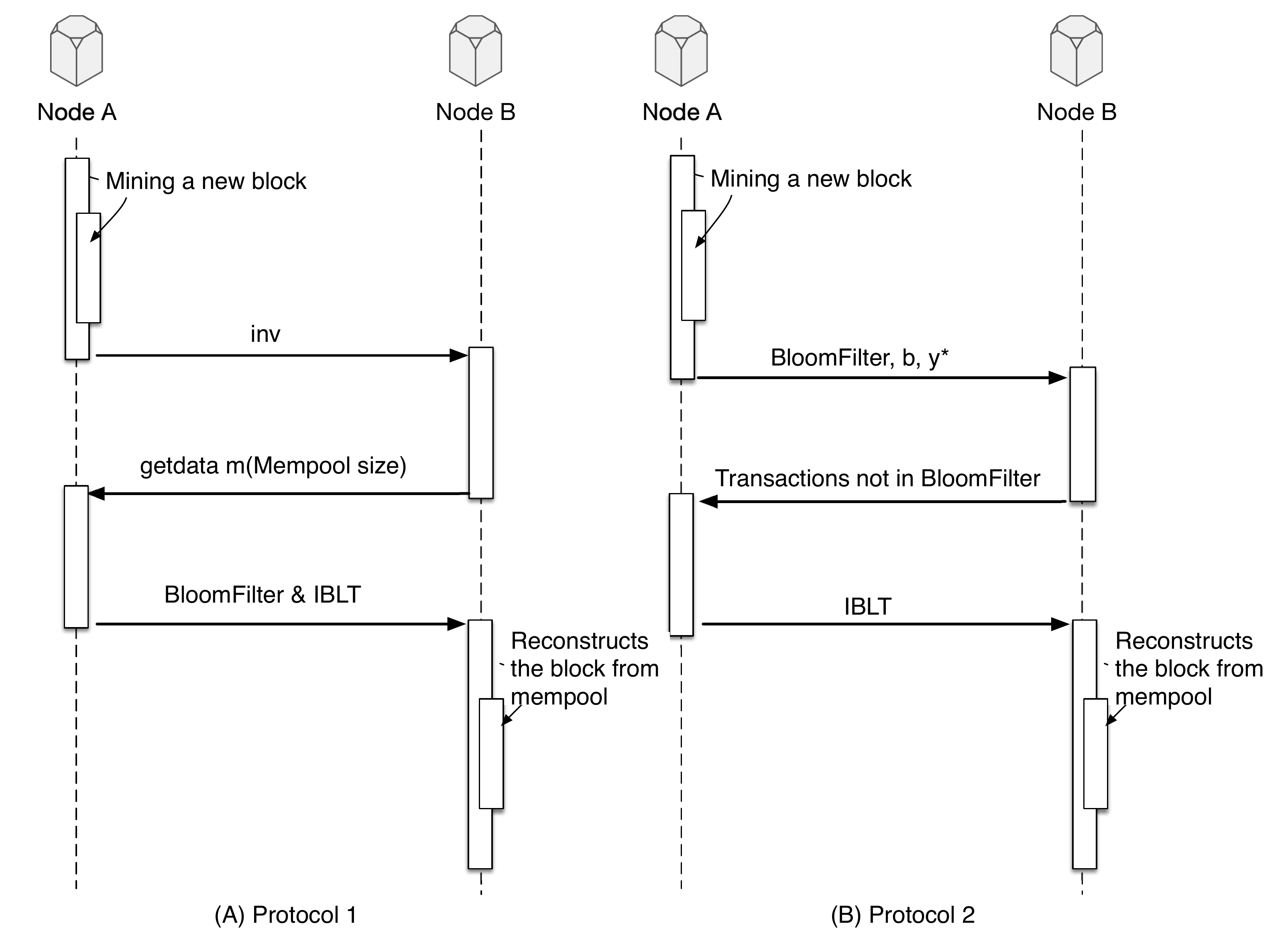}
    \caption{(A) The first protocol of Graphene synchronizes block information among different nodes through Bloom filter and IBLT, and (B) the second protocol is the compensation for the first protocol.}
    \label{Graphene}
\end{figure}

Ozisik et al.~\cite{ozisik2019graphene} proposed a novel probabilistic method called Graphene, which combines the Bloom Filter and the Invertible Bloom Lookup Table (IBLT)~\cite{goodrich2011invertible} to reduce the size of a block dramatically. Graphene meets a desired decoding rate with a high probability and is capable of reducing more bandwidth than previous approaches that use the Bloom Filter or the IBLT alone. Two protocols are designed for Graphene. The first protocol is used when the receiver’s mempool contains all the transactions in the block. In some situations, the first protocol may fail, and the second protocol must be employed. Fig.\ref{Graphene} depicts the Graphene block propagation between two nodes. 

The steps of the first protocol are shown in \reffig{Graphene}(A). Node A declares a new block of size $n$, and Node B responds with a request for the unknown block and its mempool size $m$. Node A then creates a Bloom Filter $BF$ and an IBLT $I$, respectively, using the transaction ID incorporated in the new block. The False Positive Rate (FPR) of the $BF$ is $f=\frac{a}{m-n}$, where $a$ is set to minimize the data to be transmitted. $I$ is set so that it can recover $a^*$ items, and $a^*>a$. Upon receiving the block, Node B passes the transactions from its mempool to $BF$ to create a candidate set, from which the other IBLT $I'$ is created. It then evaluates the symmetric difference through $I\Delta I'$. Finally, according to the results, it adjusts the candidate set, validates the Merkle root, and reconstructs the block. 

The second protocol is shown in \reffig{Graphene}(B), which is a remedy when the first protocol fails. In this protocol, Node A and Node B interact one more round to recover from the failure. The process is as follows. First, the candidate set is a combination of transactions in Node B's mempool and the block (i.e. true positives) and transactions in Node B's mempool but not in the block (i.e. false positives). The candidate set's size is $z$, such that $z=x+y$, where $x$ is the number of true positives and $y$ is the number of false positives. Node B calculates $x^*$ and $y^*$ such that $x^*\leq x$ and $y^*\geq y$. It also adds all transaction IDs in the candidate set to a new Bloom Filter $BF'$ with $f=\frac{b}{n-x^*}$. It then sends $BF'$, $y^*$ and $b$ to Node A, in which $b$ is the number of transactions that will be falsely recognised by Node A. After receiving $BF'$, $y^*$, and $b$, Node A can directly send the transactions in the block but not in $BF'$ to Node B. It also sends a new IBLT $L$ created from transactions in the block, from which $b+y^*$ items can be recovered. Node B creates another IBLT $L'$ from the transaction IDs already in the candidate set and newly sent by the sender. Again, it computes the symmetric difference through $L\Delta L'$ to adjust the candidate set, validates and reconstructs the block. 

A Graphene block's size under the first protocol is determined by the size of IBLT $I$ and Bloom Filter $BF$. Denoted as $G$, Graphene's block size can be calculated through \refeqs{graphene_size}, where $\tau$ is the overhead on IBLT entries, and $a$ is a parameter to minimize the size of $BF$ and $I$. \textcolor{black}{Because the accurate value of $a$ is obtained by optimization, it is difficult to calculate Graphene's block size and block capacity}. \textcolor{black}{To address this issue, a Monte Carlo simulation is conducted to evaluate the Graphene's block capacity in Section~\ref{experiment1} (Experiment 1). }

\begin{equation}
    G = n\frac{-ln(\frac{a}{m-n})}{8ln^22}+a\tau \label{graphene_size} 
\end{equation}

\subsection{XThinner} 

Similar to Compact and Xthin, XThinner \cite{XThinner} uses shorter XThinner IDs to represent the original transactions. Each ID is approximately 1.5 to 2 bytes long and encoded using transactions in the mempool by a stack-based state machine. Because its encoding mechanism is inspired by XThin, it is named after XThin. The author tested in \textcolor{black}{Bitcoin Cash} and it was reported that XThinner achieved approximately 99.6\% reduction of the block size \cite{XThinner}. XThinner is not as compact as Graphene. However, it costs less than Graphene when the receiver fails to decode the block. 

The structure of XThinner is similar to Compact. In XThinner, nodes use a flag $xtroptions$ to claim whether they can receive an XThinner block. Nodes can also specify bandwidth and latency through $xtroptions$. \textcolor{black}{In a round of XThinner's propagation}, nodes A and B are two XThinner nodes. Node B requests a newly arrived block via a $getdata$ message. Node A then replies with a $xtrblk$ which consists of a block header, a coin-based transaction, and one or more groups of shortened transaction IDs (i.e. XThinner IDs). Node B checks the validity of the block. If it is valid, Node B begins decoding the block. But once Node B fails to decode it because of missing transactions or collisions, a $xtrgettxn$ message is sent to Node A to request those missing or collided transactions. In certain situations, additional rounds of sending and receiving XThinner blocks are needed to recover from failed decoding. Finally, Nodes B checks the Merkle root of the reconstructed block. If it is matched, it begins propagating the new block. 

The length of XThinner IDs varies because it is decided according to the transaction IDs in the sender's mempool. XThinner's compression rate and block capacity cannot be directly calculated. Therefore, we conducted a Monte Carlo simulation to evaluate its block capacity (see Section~\ref{experiment1}).

\subsection{IPFS-based Storage Model}

The InterPlanetary File System (IPFS) \cite{benet2014ipfs} is a distributed data storage model. Servers connect to one another to form a P2P network to store files. Each file stored in the network is assigned a unique IPFS ID for location, generated by hashing the file using SHA2-256 algorithm.

In \cite{zheng2018innovative}, Zheng et al. proposed a centralized storage model called IPFS-based Storage Model (IPFS Model) where miners validate each transaction and store it in the IPFS. A 32-Bytes IPFS ID is returned by the IPFS and is stored in the mempool. Miners use the IPFS IDs in their mempool to construct the block and Merkle tree. When a new block is received, miners only need to validate the Merkle root using IPFS IDs and store the block in their storage. The model reduces not only the locally stored data but also data in propagation. In the Blockchain network, IPFS provides centralized storage for every node to store and share the raw data. Nodes only need to store and transmit smaller IPFS IDs. \textcolor{black}{However, it sacrifices decentralization for improving throughput.}

In IPFS Model, a transaction of 500 Bytes stored in a block is substituted by a constant-length IPFS ID of 32 Bytes. An IPFS block contains a block header and a list of IPFS IDs (including the coin-based transaction). Therefore, the block capacity is approximately equal to 32,765. \textcolor{black}{The original paper \cite{zheng2018innovative} tested the model using historical transaction data of Bitcoin. Because some early transactions were much smaller than 32 bytes, the experimental results showed a slightly low compression rate of 91.83\% compared to the theoretical rate of 93.6\%.}

\subsection{Dino}

\begin{figure}[H]
    \makeatletter
    \def\@captype{figure}
    \makeatother
    \centering
    \includegraphics[width=0.6 \textwidth]{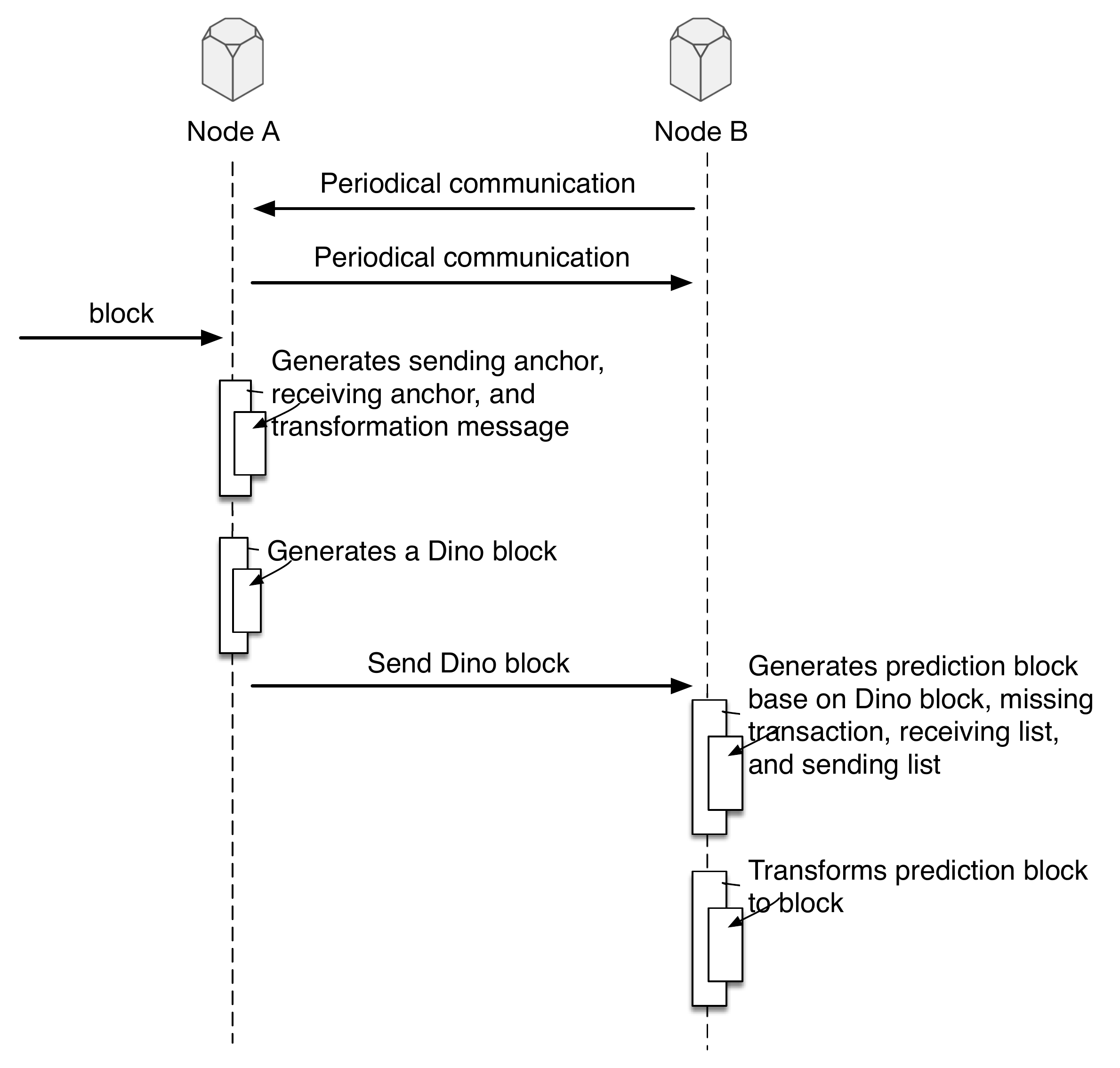}
    \caption{Illustration of the transmission mechanism in Dino.}
    \label{Dino}
\end{figure}

In most Blockchains, a miner creates a block using \refeqs{blk generation}, where $C$ is the block size, $P$ is the miner's mempool, and $F$ is the block generation algorithm. Unlike the previous five propagation protocols that directly compress the transaction data in blocks, Dino \cite{hu2022dino} transmits block construction rules. Specifically, $F$ in \refeqs{blk generation} is transmitted to reduce the network usage in propagation. The receivers reconstruct the blocks according to the rules and several missing transactions contained in the Dino block. Fig~\ref{Dino} illustrates the transmission of Dino. In Dino, the nodes' mempool contains almost all transactions packaged in the new block, and the miners tend to include transactions with higher fee rates into their candidate block. 

\begin{equation}
	BLK = F(C, P) \label{blk generation}
\end{equation}

Every node in the Dino network maintains a set of received transactions and a set of sent transactions. Nodes periodically communicate each other to synchronise their receiving and sending set. The transmission of Dino is depicted in~\reffig{Dino}. When Node A receives a new block \textsl{BLK}, it compares the block with its receiving and sending set. Transactions which are not in both sets are added to a missing set. \textcolor{black}{Node A will also locate a receiving anchor from the receiving set. Transactions beyond the anchor are considered not to be in the block. Node A also finds a sending anchor using the sending set in a similar way. Based on transactions in the missing, receiving, and sending set, Node A then constructs a predicting block \textsl{PBLK} from which it summarises the rules to transform \textsl{PBLK} into \textsl{BLK}. The rules consist of an interval set, a deleting set, and a reordering set. The interval set indicates the interval of the index of transactions belonging to \textsl{BLK} in \textsl{PBLK}. The deleting set contains transactions that should be deleted from the interval in \textsl{PBLK}. The reordering set is used to sort transactions in the interval. At last, Node A sends a Dino block containing the missing set, the receiving anchor, the sending anchor, and transformation rules to Node B. }

\textcolor{black}{Node B has its own receiving set and sending set identical to Node A's. After receiving the Dino block, Node B creates a predicting block in the same way and uses transformation rules to transform it into the original block \textsl{BLK}. A Dino block only includes the missing set, two anchors, and transformation rules. Dino's average block size increases slowly from 793 Bytes to 826 Bytes when increasing the block capacity from 2500 to 15000 according to \cite{hu2022dino}. The time complexity for the sender and the receiver in Dino is $O(nlog(n))$ and $O(n)$, respectively. Dino allows nodes to refrain from sending transaction messages periodically for better privacy conservation and lower bandwidth consumption.  }

\section{Experiments} \label{experiments}

\textcolor{black}{As described in the previous sections, each compression solution can effect the block capacity of the Blockchain. Whereas under the transaction-fee regime, the larger block capacity can lead to high fluctuation in miners' rewards due to the dynamic transaction arrivals and varying transaction amounts. To this end, in this paper, we analyze the fluctuations of miners' rewards across different compression protocols. To ensure that miner reward fluctuations are within the historical range, we estimate the maximum block size for different compression protocols. Note that, according to~\cite{croman2016scaling}, the block size should not exceed 4MB due to the Bitcoin network's current overlay and the average 10-min block interval.}

\textcolor{black}{The experiment was divided into two parts: Experiment 1 and Experiment 2. Experiment 1 adopts a Monte Carlo simulation to observe how Graphene's and XThinner's block size increases when the block capacity is increased. Experiment 2 adopts a simulation approach to evaluate the relationship between Blockchain's scalability and volatility. In both experiments, the block capacity is defined as an independent variable ranging from 1,000 to 1,000,000. The other parameter was the mempool multiplier which is a ratio of mempool size to block capacity. The mempool multiplier was calculated based on the average number of transactions in the mempool and block capacity recorded by Blockchain.com \cite{mempool_count_url} from July 2021 to July 2022. The mempool multiplier is set to 2.92, and all nodes are assumed to have the same mempool. Two sets of experiments were conducted on a Huawei Cloud Stack (HCS) Ubuntu Linux v18.04 virtual machine with a 48-core CPU at 2.5GHz, 1TB ROM, and 192GB RAM. The steps from each experiment are summarized in Table.\ref{experiments_spec}. }

\linespread{1.2}
\begin{table}[htp]
	\centering
	\caption{\textcolor{black}{Summary of the steps}}
	\label{experiments_spec}
	\begin{tabular}{p{1cm}<{\centering} |p{2cm}<{\centering} |p{8cm}}
		\toprule
		\specialrule{0em}{3pt}{1pt} 
		\textbf{Exp.} & \textbf{Compression Protocols} & \textbf{Steps} \\
		\specialrule{0em}{3pt}{1pt} 
		\hline
		\specialrule{0em}{3pt}{1pt} 
		1 & \tabincell{l}{Graphene,\\ XThinner} & \textcolor{black}{\tabincell{l}{1. Randomly generates a list of transactions.\\2. Create Graphene and XThinner blocks.\\3. Vary the block capacity, repeat the above 2 steps\\ 100 times, and collect results.}}\\
		\specialrule{0em}{3pt}{1pt}
		\hline
		\specialrule{0em}{3pt}{1pt} 
		2 & \tabincell{l}{Compact,\\ XThin,\\ Graphene,\\ XThinner,\\ IPFS Model} & \textcolor{black}{\tabincell{l}{1. Simulate transaction incorporation process based\\ on different compression protocols' block capacity. \\2. Calculate the volatility of miners' rewards for \\ different compression protocols. \\ 3. Analyze the relationship between Blockchain's\\ scalability and volatility.}}\\
		\specialrule{0em}{3pt}{1pt}
		\toprule
	\end{tabular}
\end{table}

Parameters related to both experiments are set according to the current condition of Bitcoin. Table.\ref{SimBlockSettings1} lists all the common parameters of both experiments.

\linespread{1}
\begin{table}[htp]
    \footnotesize
  \centering
  \caption{\textcolor{black}{Common parameters for both experiments}}
  \begin{threeparttable}
    \begin{tabular}{p{4cm}<{\centering} | p{4cm}<{\centering}}
    \toprule
    \textbf{Parameter} & \textbf{Value} \\
    \specialrule{0em}{3pt}{1pt}
    \hline  
    \specialrule{0em}{3pt}{1pt} 
     Mempool Multiplier & 2.92 \\
    \specialrule{0em}{3pt}{1pt}
    \hline
    \specialrule{0em}{3pt}{1pt}
    Block Capacity & [1000, 1000000] \\
    \specialrule{0em}{3pt}{1pt}
    \hline
    \specialrule{0em}{3pt}{1pt}
    \textcolor{black}{Block Header Size} & 80 Bytes \\
    \specialrule{0em}{3pt}{1pt}
    \hline
    \specialrule{0em}{3pt}{1pt}
    \textcolor{black}{Coin-based Transaction Size} & 500 Bytes \\
    \specialrule{0em}{3pt}{1pt}
    \hline
    \end{tabular}
  \end{threeparttable}
  \label{SimBlockSettings1}
\end{table}

\subsection{Experiment 1} \label{experiment1}

\textcolor{black}{Although Graphene provides a formula for its block size, it is necessary to optimize several parameters (e.g. $a$). Additionally, the formula for calculating the block size is not provided in the specification of XThinner. It is worth noting that the compression performance of Graphene and XThinner is influenced by transactions in the mempool. }
	
\textcolor{black}{In Experiment 1, we implemented a Monte Carlo simulator to evaluate the compression performance of Graphene and XThinner. By varying the block capacity, we observed the changes in block size. Each block size needs a simulation run with the following steps: (1) randomly generate a desired number of transactions which are then used to form the mempool; (2) select a portion of the transactions from the mempool and pass them to Graphene (and XThinner) to create a compressed block; (3) repeat step (1) and (2) $n$ times for each block capacity. A run generates $n$ compressed blocks with the same block capacity for Graphene (and XThinner). Based on the results, we can analyze the compression performance of Graphene and XThinner. }

\textcolor{black}{Statistically, 30 is usually accepted as the minimum number of times for a simulation. We set $n$ to 100 times considering the computational costs and simulation effectiveness. So, 100 compressed blocks with the same block capacity are generated in a simulation run.} The mempool size is calculated using the mempool multiplier and block capacity. The transaction ID is set to 8 Bytes, which is sufficient to prevent collision \cite{ozisik2019graphene}. In Graphene, a parameter called $\beta-assurance$ is used to limit its recovery rate. In the experiments, $\beta-assurance$ is set to 239/240 following the work of Ozisik et al.~\cite{ozisik2019graphene}. Table.\ref{SimBlockSettings2} lists the parameters used in Experiment 1.

\linespread{1}
\begin{table}[htp]
    \footnotesize
  \centering
  \caption{\textcolor{black}{Parameters for Experiment 1}}
  \begin{threeparttable}
    \begin{tabular}{p{4cm}<{\centering} | p{4cm}<{\centering}}
    \toprule
    \textbf{Parameter} & \textbf{Value} \\
    \specialrule{0em}{3pt}{1pt}
    \hline  
    \specialrule{0em}{3pt}{1pt} 
    Transaction ID Size & 8 Bytes \\
    \specialrule{0em}{3pt}{1pt}
    \hline
    \specialrule{0em}{3pt}{1pt}
    Trials ($n$ times) & 100 \\
    \specialrule{0em}{3pt}{1pt}
    \hline
    \specialrule{0em}{3pt}{1pt}
    $\beta-assurance$ & 239/240 \\
    \specialrule{0em}{3pt}{1pt}
    \hline
    \end{tabular}
  \end{threeparttable}
  \label{SimBlockSettings2}
\end{table}

\begin{figure}[htbp]
    \centering
    \subfloat{
        \includegraphics[width=0.3\textwidth]{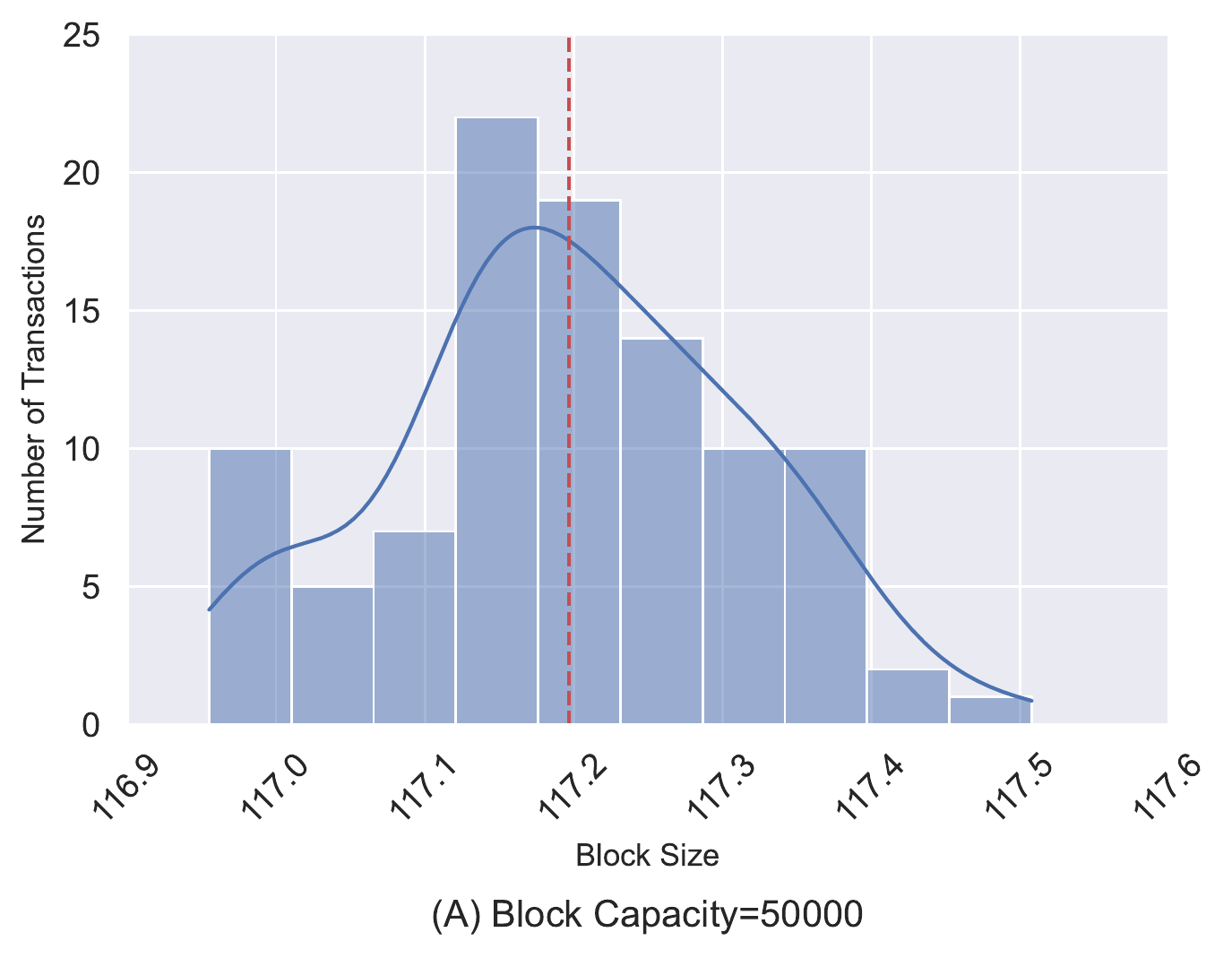}
    }
    \subfloat{
        \includegraphics[width=0.3\textwidth]{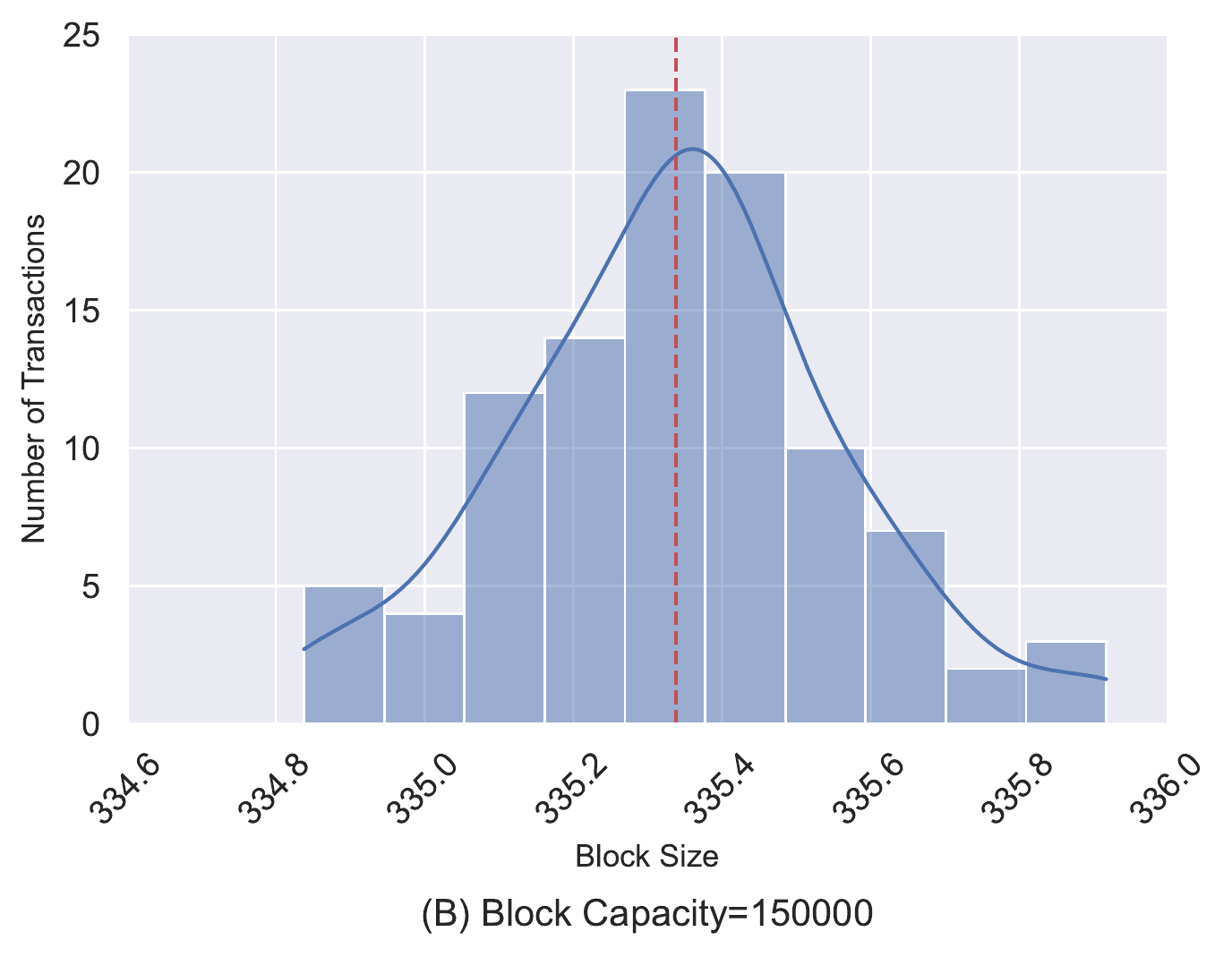}
    }
    \subfloat{
        \includegraphics[width=0.3\textwidth]{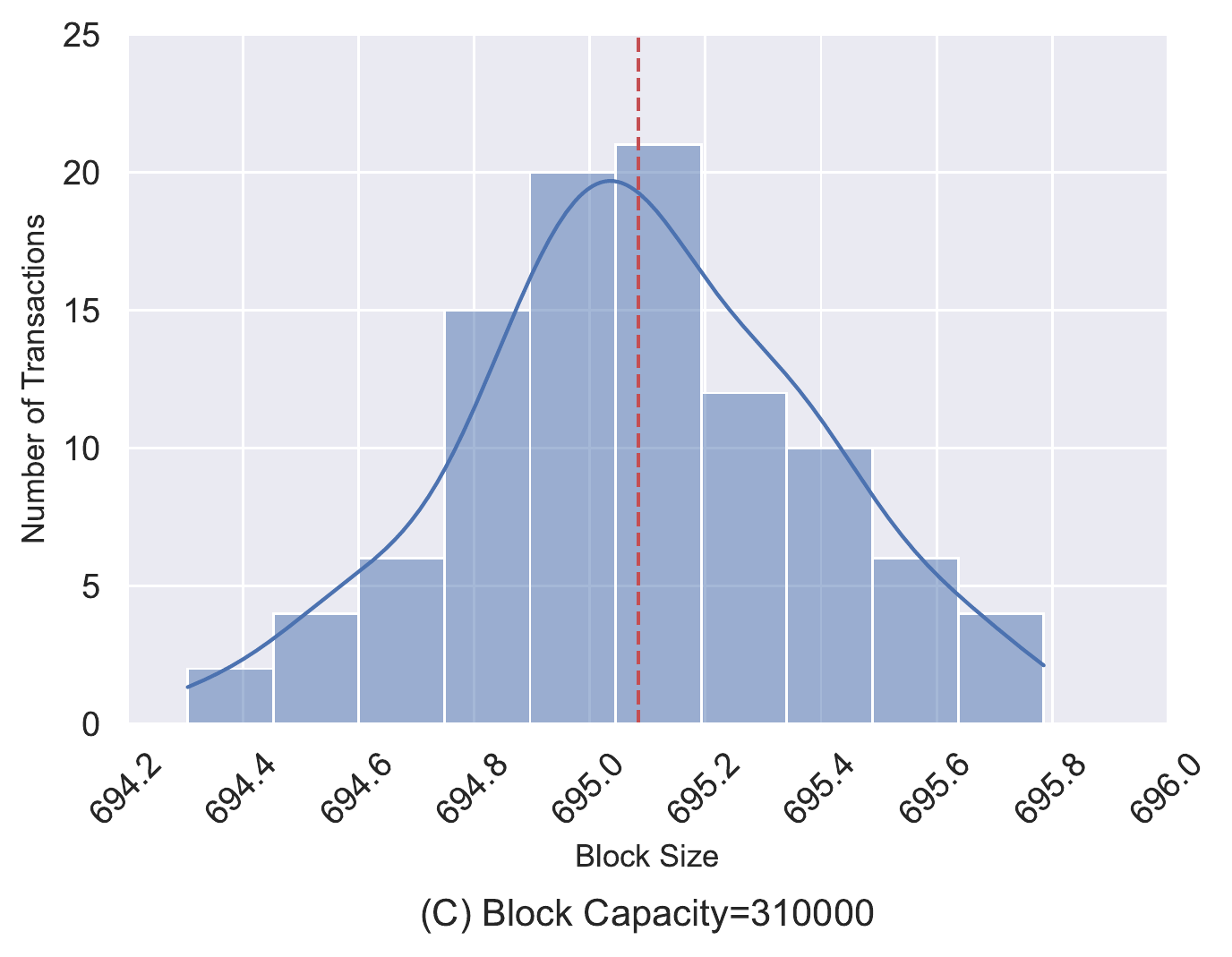}
    }\hspace{0.001\textwidth} \\[-2ex]
    \subfloat{
        \includegraphics[width=0.3\textwidth]{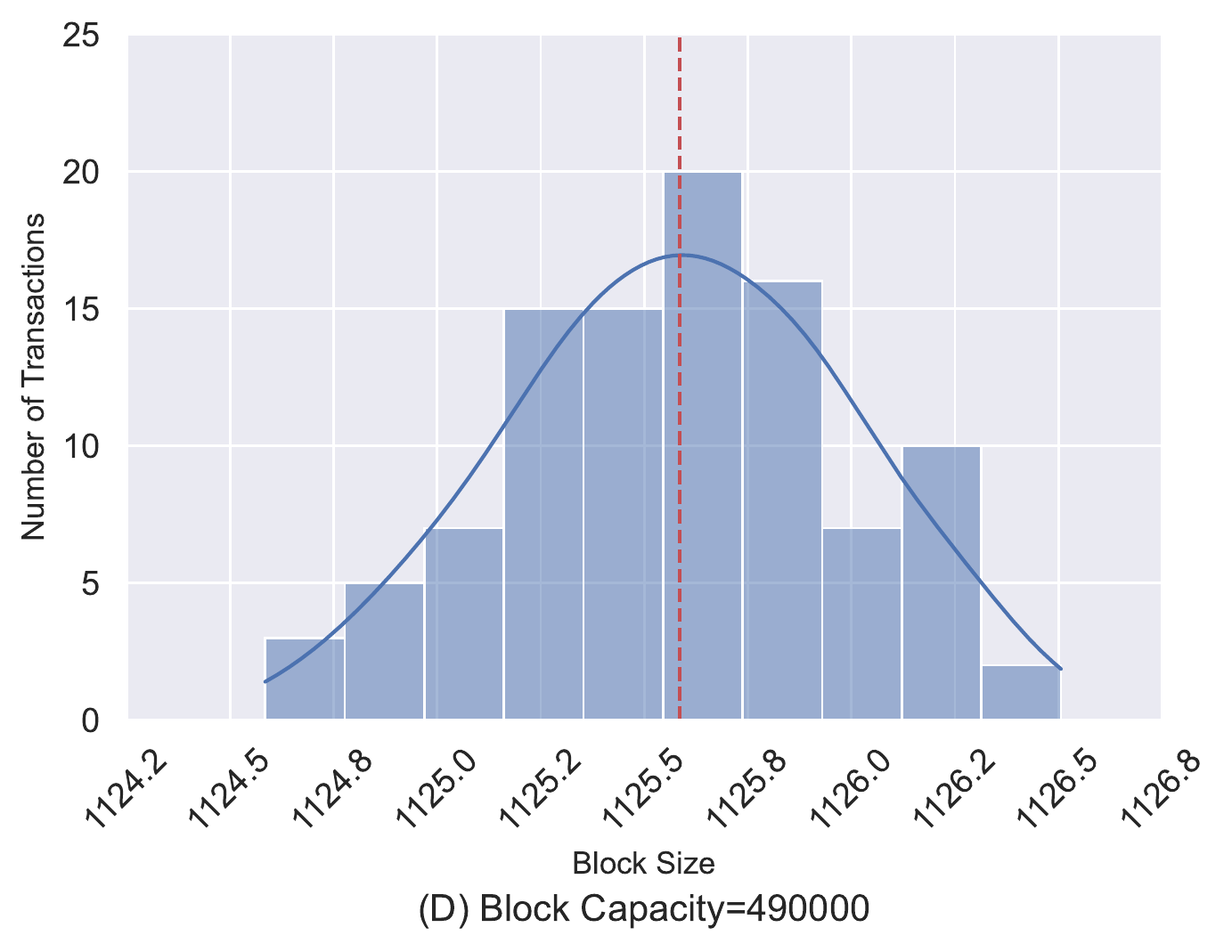}
    }
    \subfloat{
        \includegraphics[width=0.3\textwidth]{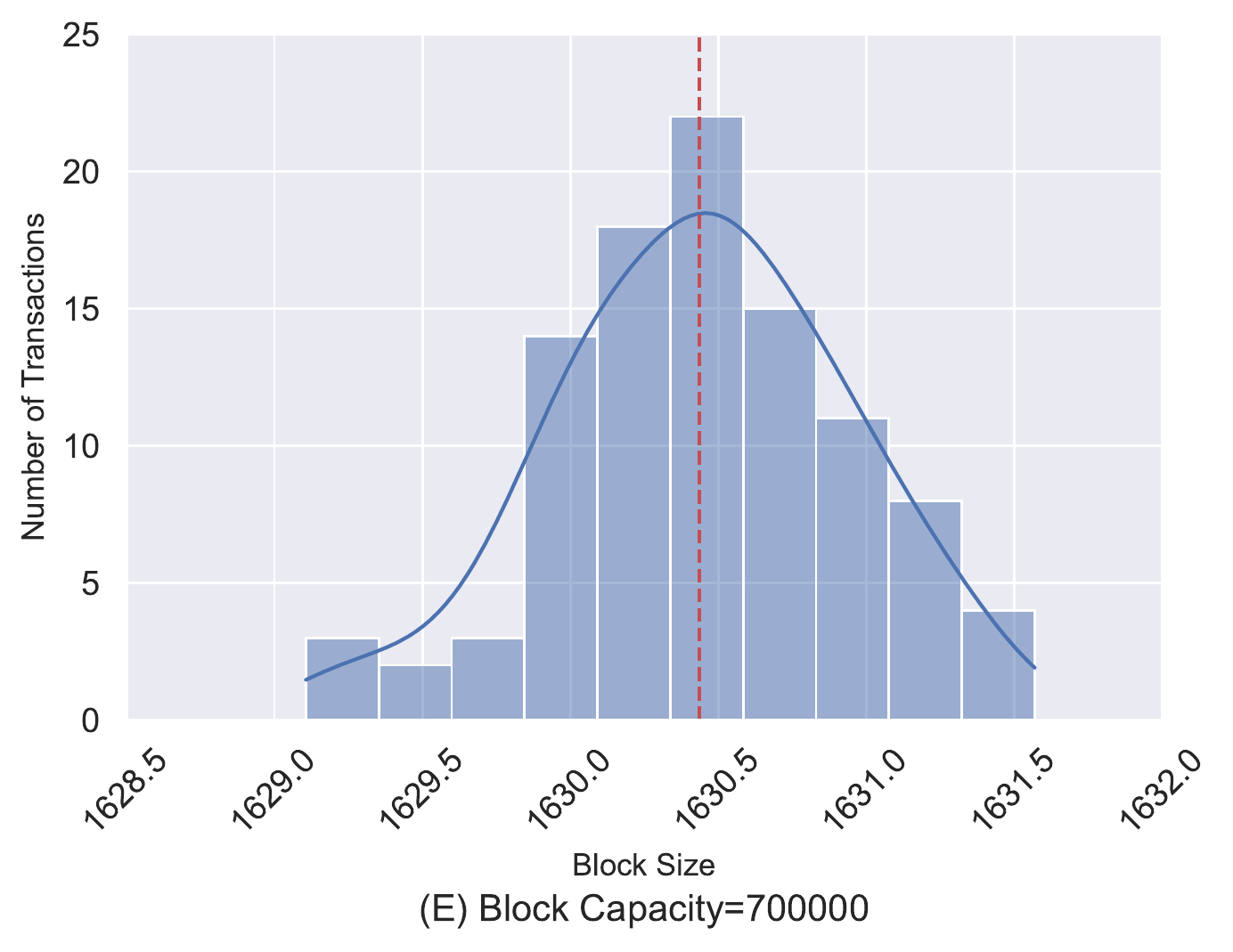}
    }
    \subfloat{
        \includegraphics[width=0.3\textwidth]{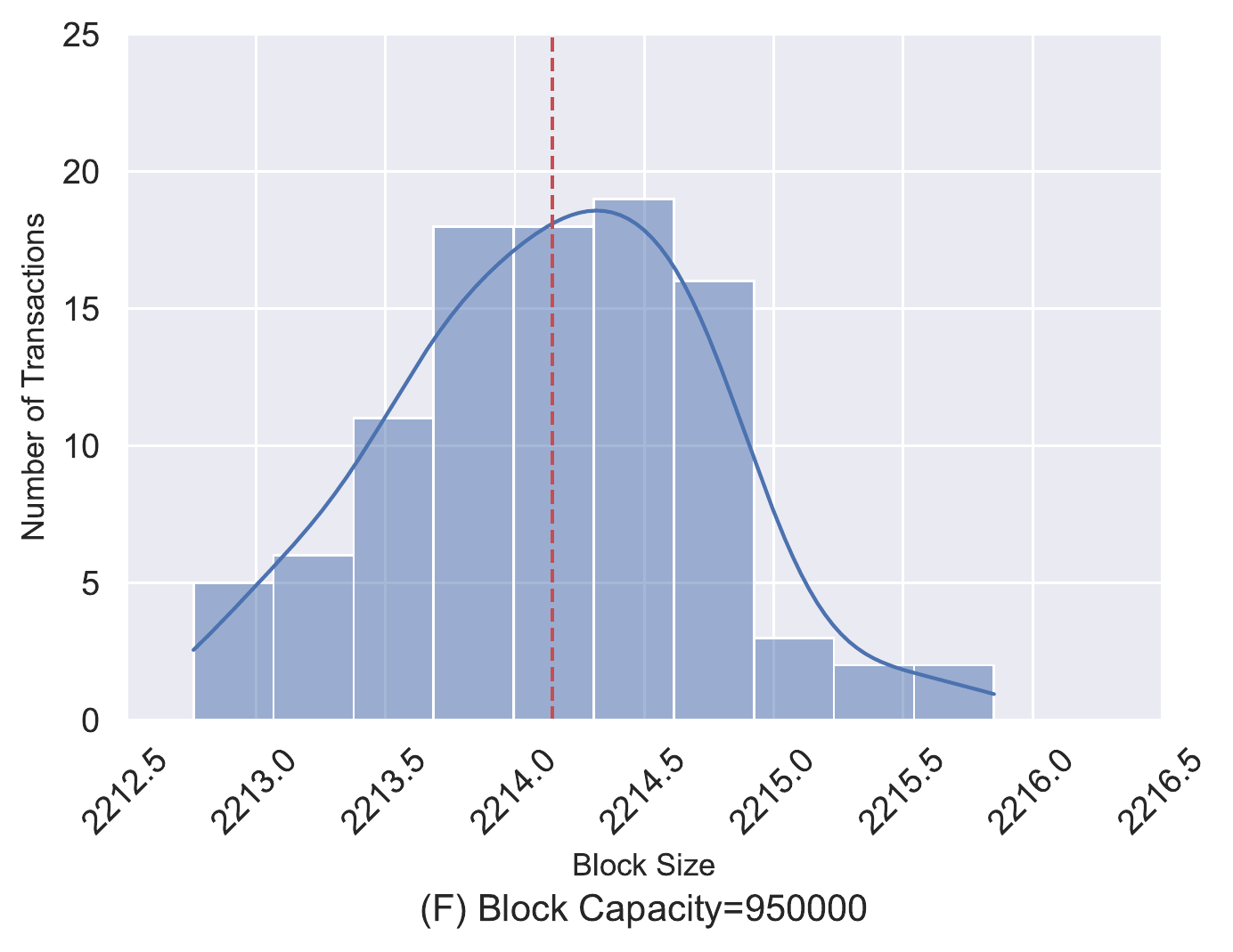}
    }
    \caption{Histogram of XThinner's block size (KB). Each subfigure plots the block size distribution under a specific block capacity. The size almost follows a normal distribution.}
    \label{xthinner_hist}
\end{figure}

\textcolor{black}{In Experiment 1, the simulator generates a new mempool every time it is executed. When the block capacity remains the same, the 100 Graphene blocks have the same size. It implies that Graphene's block size is unaffected by different mempools. So we do not discuss Graphene in this paragraph. In contrast, the size of XThinner blocks in the same simulation run varies slightly since it depends on transactions in the mempool, that is, the raw data of the transactions. To illustrate this point, we draw a group of histograms of the compressed size of XThinner for 6 different block capacities. The 6 block capacities are uniformly selected from the range of block capacity, and the block size distribution under other block capacities is similar to the selected subfigures. As shown in \reffig{xthinner_hist}, we plot a histogram based on the 100 XThinner blocks for each block capacity. In each histogram, the x-axis is the size of compressed blocks. The y-axis is the number of blocks. The red dotted line represents the mean value, while the blue curve is the kernel density estimation. We can observe that the block size almost follows a normal distribution with a small range (within 3 KB). Therefore, we use the mean of the 100 blocks to represent XThinner's block size for each block capacity.  } 

\begin{figure}[H]
    \makeatletter
    \def\@captype{figure}
    \makeatother
    \centering
    \includegraphics[width=0.6 \textwidth]{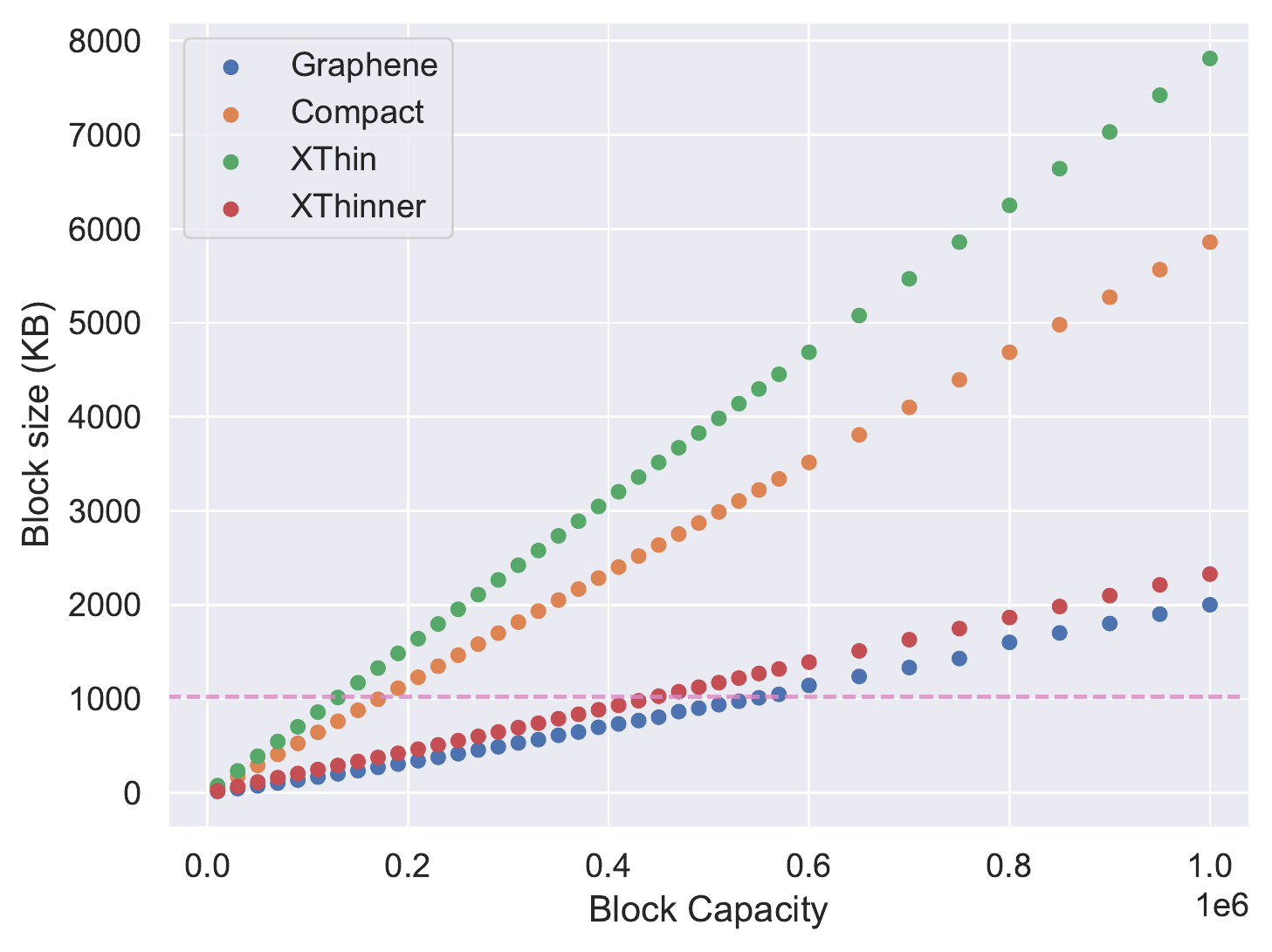}
    \caption{Size of four protocols' blocks as the block capacity increases. The pink dotted line represents the block size of 1024 KB.}
    \label{graphene_capacity}
\end{figure}


\textcolor{black}{Through the Monte Carlo simulation, the block size for different block capacities of Graphene and XThinner is obtained. The block size is then used in Experiment 2. Next, we compare the compression performance of four protocols: Compact, XThin, Graphene, and XThinner. We can calculate the block size of Compact, XThin, and IPFS Model based on their compression rates provided in the original specifications. But IPFS Model's block size increases fast as the block capacity grows. So, we only included Compact and XThin in the comparison. According to the original paper, nodes under Dino protocol relay blocks containing the block generation rules. The block size is almost constant, regardless of how many transactions will be transmitted. During the average 10-min block interval, Dino nodes frequently communicate with each other and synchronize the transactions to ensure efficient transmission of the block generation rules.} Thus, the relation between block capacity and block size cannot be calculated directly based on a formula or obtained by a Monte Carlo simulation. Besides, the source code of Dino is not publicly available. Therefore, Dino was not included in Experiment 1 and Experiment 2.

\reffig{graphene_capacity} shows how the block size changes for each protocol as the block capacity increases. In \reffig{graphene_capacity}, the pink dotted line represents a block size of 1024 KB. From \reffig{graphene_capacity}, we can observe that the size of Graphene blocks is smaller than XThinner blocks. However, block sizes of Graphene and XThinner grow much slower than XThin and Compact blocks. The size of Graphene and XThinner blocks increases faster when the block capacity is increased, which implies their compression rates decrease slightly with the increasing block capacity. 


\subsection{Experiment 2} 

\textcolor{black}{In this section, we illustrate Experiment 2 and analyze its results.} SimBlock \cite{8751431} was employed to evaluate the miners' behaviour under different block capacities to analyze the relationship between the Blockchain's volatility and scalability. The consensus protocol used in SimBlock is Proof-of-Work (PoW), but other consensus protocols such as Proof-of-Burn (PoB) and Proof-of-Stake (PoS) can also be employed. During the initialization, SimBlock tops up the mempool with the Bitcoin historical transaction data. All nodes in the Bitcoin network communicate and share the transactions in their mempool. During the simulation, transactions in the mempool are packaged into a block every 10 minutes. To observe the volatility and scalability caused by different block capacities, we implemented a Python control program that runs SimBlock and varies the block capacity automatically. After the simulation, results are collected and used to calculate the volatility and throughput of a specific block capacity. Table.\ref{SimBlockSettings3} lists the parameters used in Experiment 2. 

\linespread{1}
\begin{table}[htp]
  \centering
  \caption{\textcolor{black}{Parameter for Experiment 2}}
  \begin{threeparttable}
    \begin{tabular}{p{4cm}<{\centering} | p{4cm}<{\centering}}
    \toprule
    \textbf{Parameter} & \textbf{Value} \\
    \specialrule{0em}{3pt}{1pt}
    \hline  
    \specialrule{0em}{3pt}{1pt} 
    Block Size & 1050000 Bytes \\
    \specialrule{0em}{3pt}{1pt}
    \hline
    \specialrule{0em}{3pt}{1pt}
    Transaction Size & 500 Bytes\\
    \specialrule{0em}{3pt}{1pt}
    \hline
    \specialrule{0em}{3pt}{1pt}
    Transaction Fee Percentage & 0.2\% \\
    \specialrule{0em}{3pt}{1pt}
    \hline
    \toprule
    \end{tabular}
  \end{threeparttable}
  \label{SimBlockSettings3}
\end{table}

\subsubsection{Datasets}

\begin{figure}[tbp]
    \centering
    \subfloat{
        \label{value:a}
        \includegraphics[width=0.6\textwidth]{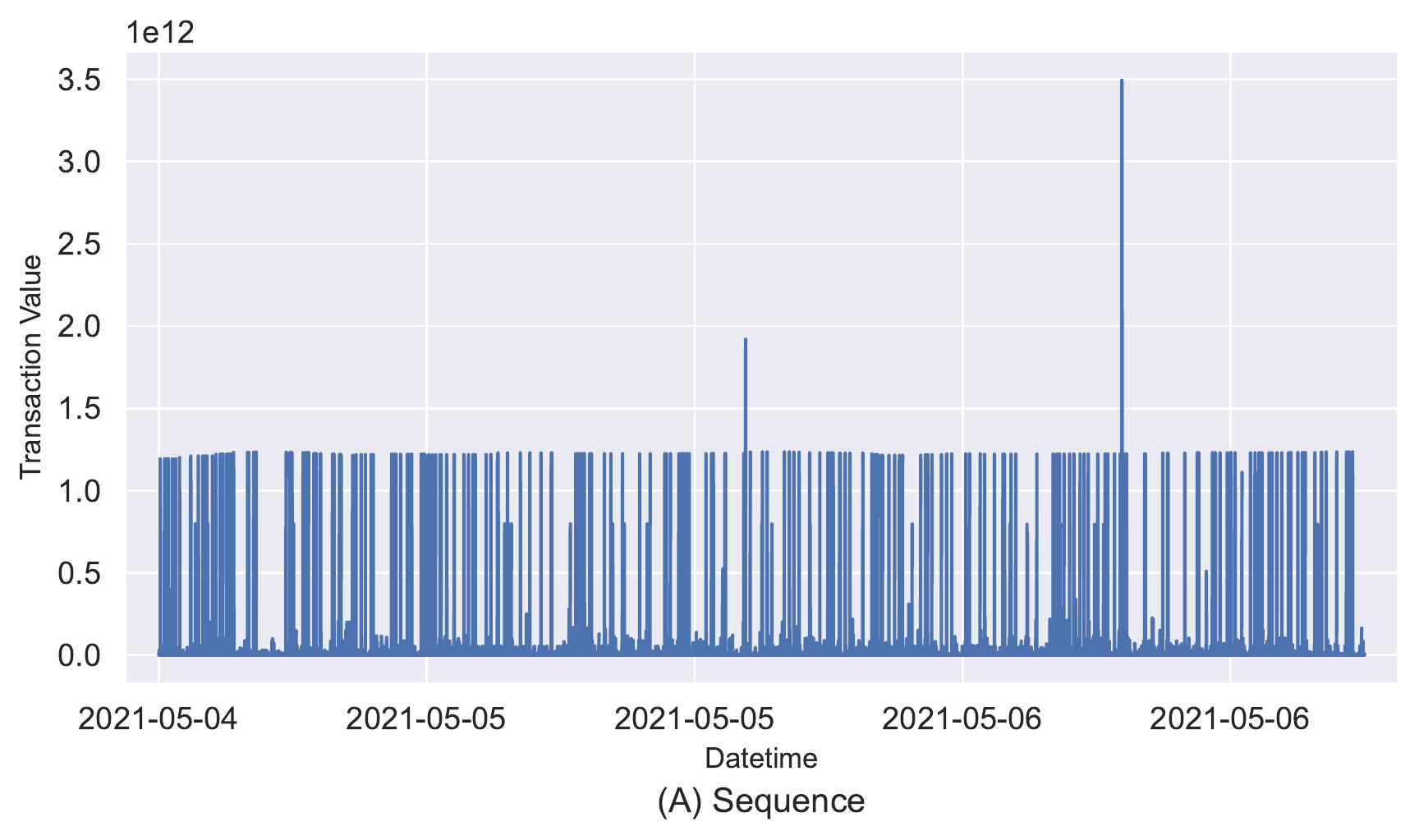}
    }\\[-1ex]
    \subfloat{
        \label{value:b}
        \includegraphics[width=0.6\textwidth]{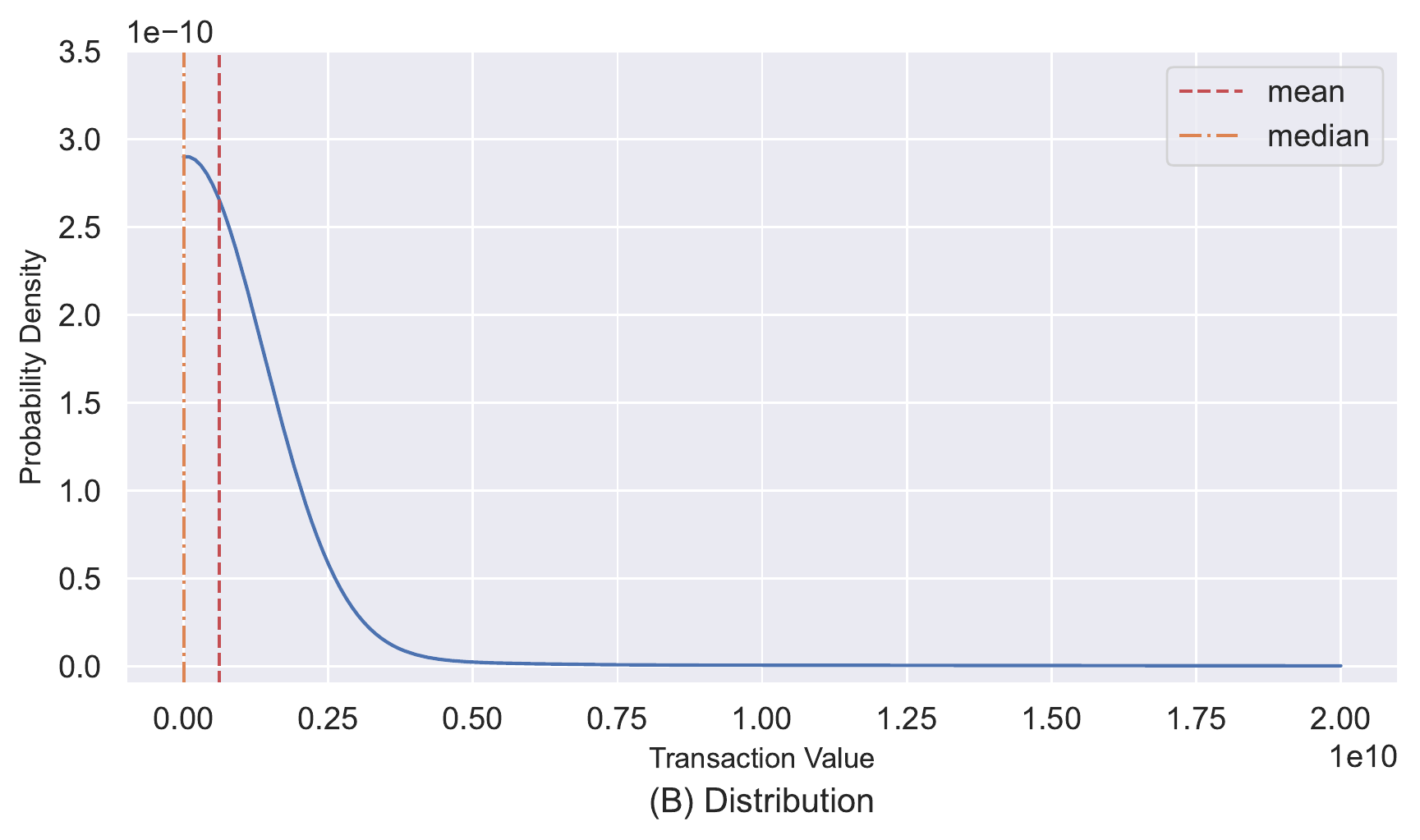}
    }
    \caption{(A) Block values in time sequence (B) The distribution of transaction values and the red and orange lines are mean and median values, respectively.}
    \label{bit_value}
\end{figure}

\textcolor{black}{Experiment 2 is based on Bitcoin historical transaction data from May 1 to November 1, 2021, consisting of 40,000,000 transactions. \reffig{bit_value}(A) shows 900,000 transaction input values from May 4 to May 6 in chronological sequence. The x-axis is the date, while the y-axis is the corresponding transaction value. \reffig{bit_value}(A) depicts uneven input values with high volatility. Meanwhile, \reffig{bit_value}(B) shows the distribution of input values. Because the tail in~\reffig{bit_value}(B) is too long, the plot is truncated at $2^{10}$. The x-axis is the input value, while the y-axis is the probability density. \reffig{bit_value}(B) shows that most transactions have low values, as evident from the high peak on the left. But the long tail behind the peak indicates the presence of extreme values in trading. The red dotted line in the plot is the mean value of 612,542,247, and the orange dotted line is the median of 1,782,395. \reffig{bit_value}(A) and (B) also reveal that extreme input values are rare in our dataset. Transactions with an input value above $0.5\times10^{12}$ only take a portion of 0.0463\%. However, these extremely high-value transactions have a major impact on the  miners' rewards. }

\subsubsection{Benchmarks}
In experiment 2, we used Historical Volatility (HV) to measure the volatility of miners' rewards and Transactions Per Second (TPS) to measure the throughput of the Blockchain. 
\begin{itemize}

\item \textbf{Historical Volatility (HV):} HV is widely used in analyzing the stock market. It reflects the price fluctuation of the asset in the past period of time. In our experiments, HV is used to measure the fluctuation of miners' rewards. HV is a critical indicator in the transaction-fee regime for ensuring the Blockchain's sustainability and security. We denote $I_m$ as the revenue of $m^{th}$ block, $I_{m-1}$ as the revenue of $(m-1)^{th}$ block, and $\Bar{I}$ as the average revenue of all $n$ blocks. HV can be calculated as follows. 

\begin{equation}
    HV = \sqrt{\frac{\sum^m_{i=1}(ln(I_i/I_{i-1})-\Bar{I})^2}{n-1}} \label{volatility}
\end{equation}

According to \refeqs{volatility}, we can calculate the yearly HV of miners' rewards in the past 10 years from 2012 to 2021 using Bitcoin’s daily miners' revenue provided by NASDAQ \cite{nasdaq}. Our simulation results are compared to the calculation results used as real-world volatility references. The real-world volatility values are listed in Table.\ref{HistoricalVolatility}, and they are plotted in \reffig{real_HV}. From \reffig{real_HV}, we can observe that the minimum and maximum HV are 0.037647 and 0.238111. These values were recorded in 2019 and 2012. Note that HV in recent years has been decreasing. Over the past 6 years, HV has remained steady below 0.075. Under the fee regime, if HV had soared to where it was 10 years ago or even higher, that would be an eight-fold increase. High volatility would significantly increase the probability of deviant mining strategies. 

\linespread{1.2}
\begin{table}[htp] 
  \footnotesize
  \centering
  \caption{\textcolor{black}{Historical Volatility of miners' rewards from 2012 to 2021}}
  \begin{threeparttable}
    \begin{tabular}{p{1.5cm}<{\centering}|p{3.5cm}<{\centering}|p{1.5cm}<{\centering}|p{3.5cm}<{\centering}}
      \toprule
      \specialrule{0.00em}{3pt}{1pt} 
      \textbf{Year} & \textbf{Historical Volatility} & \textbf{Year} & \textbf{Historical Volatility}\\
      \specialrule{0.00em}{3pt}{1pt} 
      \hline
      \specialrule{0.00em}{3pt}{1pt} 
      \textbf{2012} & \textbf{0.238111} & 2013 & 0.200857\\
      \specialrule{0.00em}{3pt}{1pt} 
      \hline
      \specialrule{0.00em}{3pt}{1pt} 
      2014 & 0.218010  & 2015 & 0.180948\\
      \specialrule{0.00em}{3pt}{1pt} 
      \hline
      \specialrule{0.00em}{3pt}{1pt} 
      2016 & 0.073051  & 2017 & 0.063965\\
      \specialrule{0.00em}{3pt}{1pt}  
      \hline
      \specialrule{0.00em}{3pt}{1pt}  
      2018 & 0.045616 & \textbf{2019} & \textbf{0.037647}\\
      \specialrule{0.00em}{3pt}{1pt}  
      \hline
      \specialrule{0.00em}{3pt}{1pt}  
      2020 & 0.059485  & 2021 & 0.044932 \\
      \specialrule{0.00em}{3pt}{1pt}  
      \toprule
    \end{tabular}
  \end{threeparttable}
  \label{HistoricalVolatility}
\end{table}

\begin{figure}[htp]
    \makeatletter
    \def\@captype{figure}
    \makeatother
    \centering
    \includegraphics[width=0.6 \textwidth]{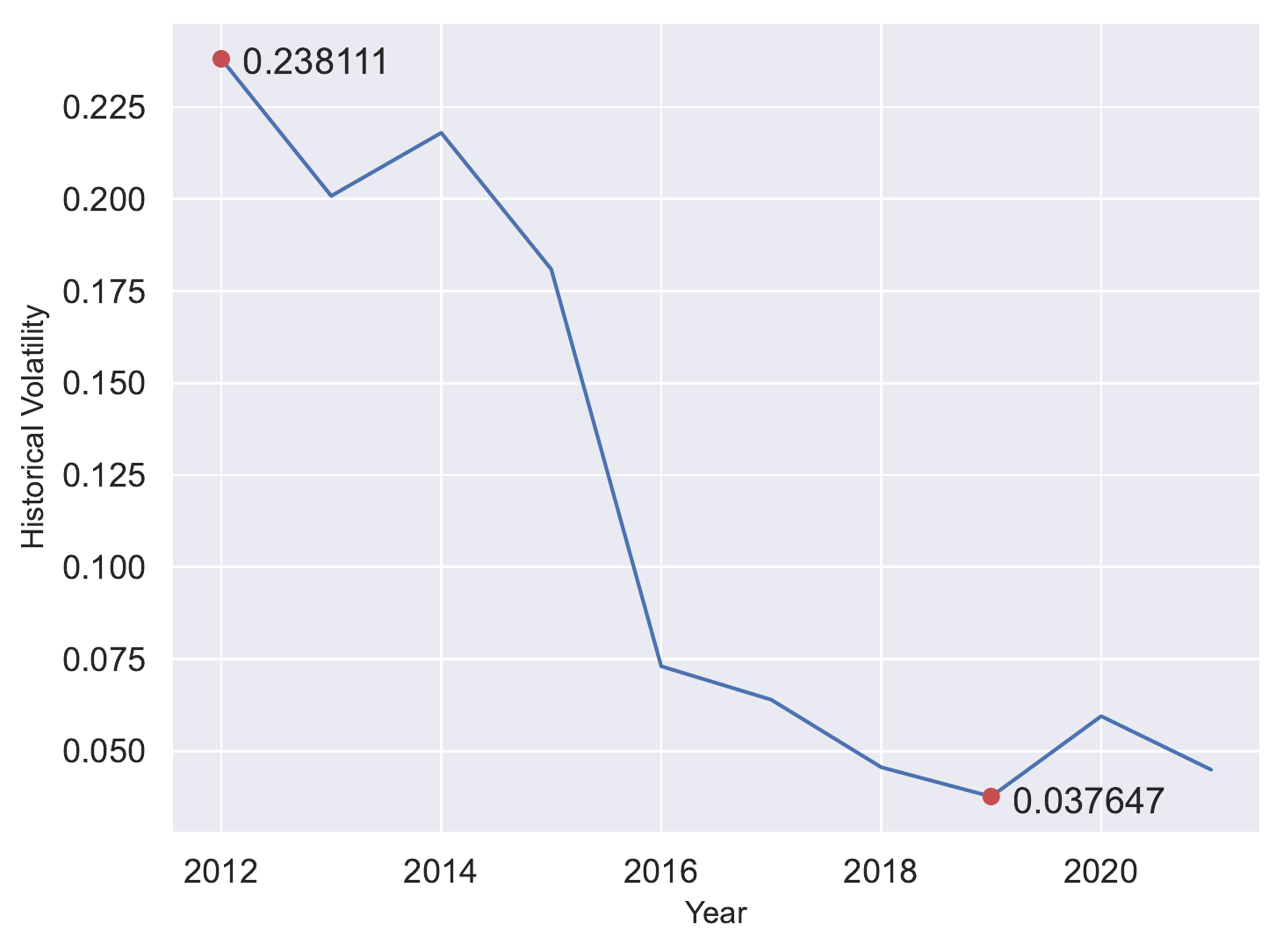}
    \caption{Historical Volatility from 2012 to 2021 with a maximum of 0.238 in 2012 and a minimum of 0.038 in 2019.}
    \label{real_HV}
\end{figure}

\item \textcolor{black}{\textbf{Transactions Per Second (TPS)}: TPS is an important metric in measuring the throughput of Blockchain. Given the block size and block time, TPS can be calculated using \refeqs{TPS}. Currently, the throughput of Bitcoin is approximately in the range of 3.3 to 7 TPS, depending on the size of transactions incorporated. For comparison, VISA's maximum TPS is about 1700, and the throughput per 10 minutes is approximately 1,020,000. In this experiment, we set the range of the block capacity variable between 1,000 to 1,000,000 to investigate how the volatility of miners' rewards changes when the Blockchain's throughput approaches VISA's TPS.} 

\end{itemize}

\begin{equation}
    TPS = \frac{Block\ Size}{Block\ Time} \label{TPS}
\end{equation}


\subsubsection{Experimental Results}
\begin{figure}[htp]
    \makeatletter
    \def\@captype{figure}
    \makeatother
    \centering
    \includegraphics[width=0.6 \textwidth]{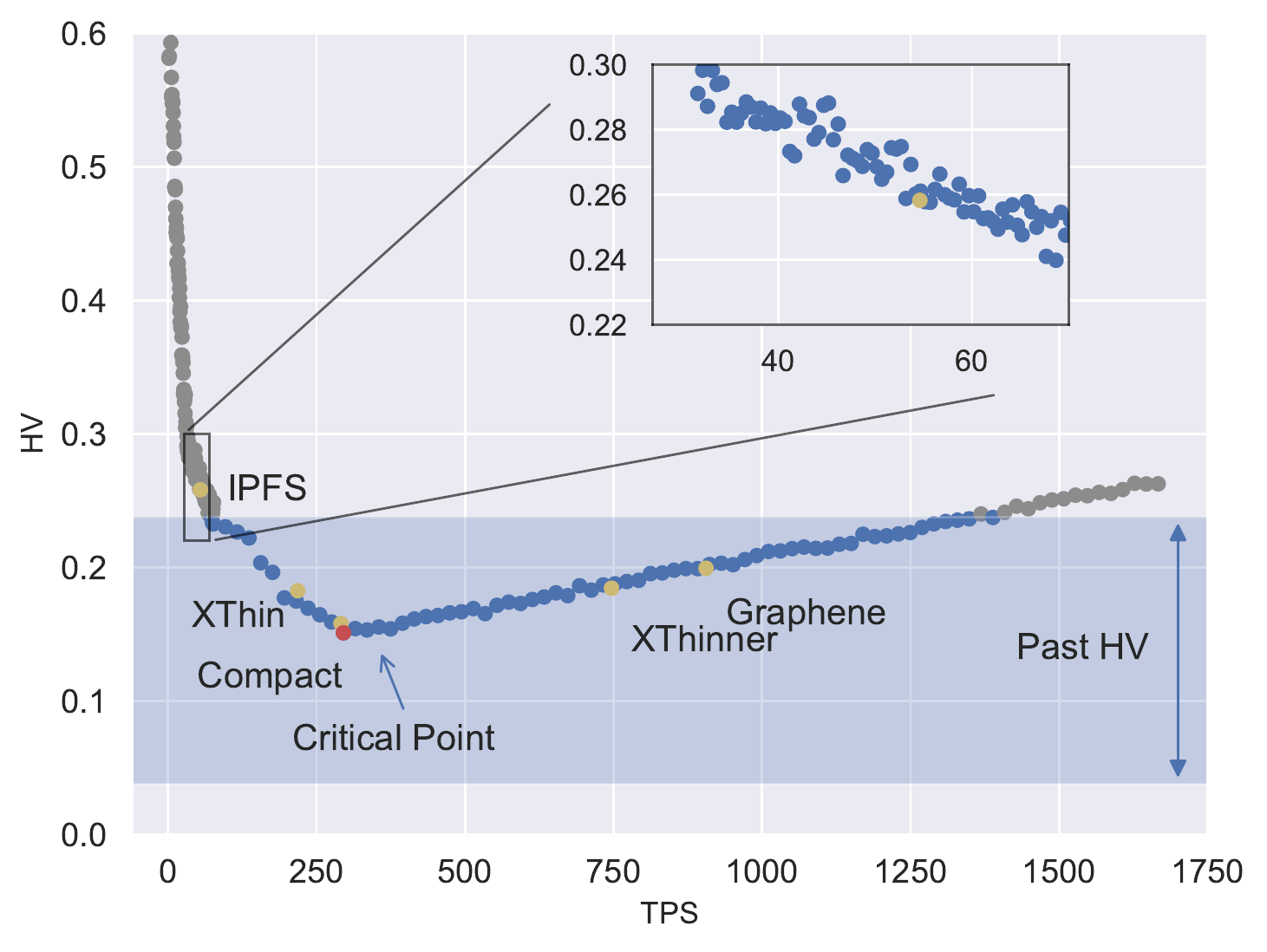}
    \caption{Relationship between TPS and HV with Past HVs (blue shade) as a reference. Five compression protocols are marked in yellow, and the critical point is in red.}
    \label{Volatility}
\end{figure}

In Experiment 2, the behaviour of packaging transactions into blocks of different block sizes was simulated using SimBlock. Based on the simulation results, the HV of miners' rewards was calculated. In the experiment, we use throughput (TPS) to represent block capacity because they are linearly dependent. \reffig{Volatility} shows the relationship between TPS and HV, where points form a curve with a critical point. The historical range of real-world HV in the past decade is depicted in blue colour shadow. Points outside the real-world HV are marked in grey and the other points are in deep blue. \textcolor{black}{The highest HV in simulation is 0.58 with a TPS of 1.67. The lowest HV is 0.15 with a TPS of 295.29 which can be considered as the critical point (turning point). Between TPS around 30 to 70, although the overall trend is decreasing, the HV of the points fluctuates in a small range. To illustrate this condition, we draw a detailed subplot for TPS between 30 to 70 in \reffig{Volatility}.} 
	
\textcolor{black}{In \reffig{Volatility}, we can observe a critical point. When the throughput of the Bitcoin network is below the critical point, the volatility sharply decreases as the throughput increases. However, the volatility gradually increases when the throughput is above the critical point. In our experiments, we find that the location of the critical point can vary due to the conditions such as mempool size, transaction fee percentage, and the input value distribution.} \textcolor{black}{According to the real-world HV, we define two terms: underscaling and overscaling. Underscaling refers to the condition where the Blockchain is not scaled sufficiently, and its throughput is lower than the critical point. And overscaling is the condition where the Blockchain is scaled excessively, and its throughput is higher than the critical point. They both result in higher HV than the real-world range.}

\textcolor{black}{Through Experiment 1, we obtained the block capacity of Graphene and XThinner. We also calculated the block capacity of Compact, XThin and IPFS Model. And consequently, we can simulate the corresponding HV of their block capacities. In \reffig{Volatility}, we highlight all five protocols in yellow. Their locations in the figure present the HV and TPS they will reach when the block size is fixed at 1 MB. } 
	

\subsubsection{Observation on the critical point}
In this subsection, we further analyze the critical point and security issues thereafter. \textcolor{black}{Higher throughput is beneficial for miners to increase their rewards and decrease the volatility of the rewards. However, when the Blockchain is overscaled, higher throughput may also worsen the stability of block revenue.} 

Before reaching the critical point, a block can only incorporate a limited number of transactions, resulting in higher revenue volatility than the critical point due to the significant fee differences between transactions. \textcolor{black}{In addition, because of underscaling and the randomness in transaction arrivals, miners' rewards and the volatility also present fluctuation when TPS is between 30 and 70.} As the block capacity grows, more transactions can be packaged into a block. High-value transactions are more evenly distributed in the blocks. \textcolor{black}{The fluctuation of HV in a small range still exists, but as the TPS increase from 70, higer total transaction fees in a block make the fluctuation tiny.} However, miners' rewards become unbalanced when the block capacity exceeds the critical point. This situation happens when more high-value or low-value transactions are incorporated into the same block due to the time-varying nature of transaction fees. 

When transaction fees become the miners' primary revenue source, stable rewards are critical to maintain their income expectation and prevent deviant mining behaviours \cite{9592512}. We adopted the HV in the past 10 years as a real-world reference for our experiment (\reffig{real_HV}). \textcolor{black}{If the Blockchain is underscaled or overscaled, the volatility of miners' rewards will exceed the real-world HV.} Such deviation from the persistently low volatility is detrimental to the Blockchain's security as the real-world HV has been less than 0.075 for about 6 years. High-value blocks become attractive to miners to steal the rewards therein. If the miners' rewards fluctuate dramatically, they cannot make reasonable expectations about their rewards. Even if miners find a block, the potential block reward may not cover their power costs. As a result, deviant mining threats such as Mining Gap may occur. The number of miners will decrease for this Blockchain, compromising its security. In addition, if all the miners use the default compliance strategy, they will mine at the top of the longest chain and authorize all available transactions. Miners know their rewards and may wait for profitable transactions in the mempool before completing a block. Such behaviour reduces mining time. Obviously, it will negatively impact the security of the Blockchain since the effective hashing capacity in the network will drop further. As a result, malicious miners will be more likely to fork the Blockchain~\cite{2978408}.

\subsubsection{Observation on compression protocols} \textcolor{black}{In Table.\ref{EvaluationTable}, we summarised the six selected compression protocols from the perspectives of their compression rate (Byes Per Transaction), block capacity, throughput (TPS), HV, acceptable block size (Acceptable Size) and maximum block size (Max size).}

\textcolor{black}{Our results in \reffig{Volatility} reveal a scaling limit which is around 1350 TPS to avoid overscaling. Regardless of how high the compression rate of a solution is, the number of transactions packaged in a block is not recommended to exceed the limit. Based on the highest TPS of 1350, the acceptable block size of each protocol is calculated and reported in the Acceptable Size column (in Kilobytes). However, according to Croman et al. \cite{croman2016scaling}, the maximum block size should be limited to 4 MB, considering the network issues and block interval. Therefore, the acceptable block sizes of Compact, XThin, and IPFS Model are found to have exceeded the upper bound (4 MB). In the last column, we propose the recommended maximum block size for each protocol while taking into account the upper bound constraint.}

\linespread{0.9}
\begin{table}[htp]
    \footnotesize
	\centering
	\caption{\textcolor{black}{On-chain scaling solutions comparison}}
	\label{EvaluationTable}
	\begin{tabular}{p{1.25cm}<{\centering} |p{2cm}<{\centering} |p{1.3cm}<{\centering} |p{0.9cm}<{\centering} |p{1cm}<{\centering} |p{1.3cm}<{\centering} |p{1cm}<{\centering}}
		\toprule%
		\textbf{Protocol} & \textbf{Bytes Per Transaction} & \textbf{Block Capacity} & \textbf{TPS} & \textbf{HV} & \textbf{Accept-able Size} & \textbf{Max Size}\\ [1ex]
		\midrule
		Compact & 6 & 174,663 & 291.10 & 0.1581 & 4746.09 & 4096 \\ 
		\specialrule{0em}{3pt}{1pt}
		\hline
		\specialrule{0em}{3pt}{1pt} 
		XThin & 8 & 130,999 & 218.33 & 0.1827 & 6328.15 & 4096 \\ 
		\specialrule{0em}{3pt}{1pt}
		\hline
		\specialrule{0em}{3pt}{1pt}  
		Graphene & 1.93 (varied) & 543,379 & 905.63 & 0.1995 & 1762.22 & 1812.34 \\ 
		\specialrule{0em}{3pt}{1pt}  
		\hline
		\specialrule{0em}{3pt}{1pt}  
		XThinner & 2.34 (varied) & 447,913 & 746.52 & 0.1846 & 1890.09 & 1890.09 \\ 
		\specialrule{0em}{3pt}{1pt}
		\hline
		\specialrule{0em}{3pt}{1pt} 
		IPFS & 32 & 32,765 & 54.61 & 0.2583 & 25312.50 & 4096 \\ 
		\specialrule{0em}{3pt}{1pt}
		\hline
		\specialrule{0em}{3pt}{1pt}  
		Dino & ca.800 (a block) & NA & NA & NA & NA & NA\\
		\specialrule{0em}{3pt}{1pt}
        \toprule
	\end{tabular}
	\begin{tablenotes}
		\item NA - Not Applicable, ``-" - indicates Not available.\\
	\end{tablenotes}
\end{table}

\textcolor{black}{Compact has the lowest HV among all the protocols, while IPFS Model has the highest. IPFS Model is the only model that results in underscaling. Graphene and XThinner have higher compression rates than Compact and XThin, but they also exhibit higher volatilities. It is sufficient to avoid underscaling and the high volatility that appears in underscaling condition if the existing scaling solutions are applied.} 

It appears that higher compression rates (lower BPT) are not desired in the future development of compression protocols due to the scaling limit. However, in the context of networking, for blocks that exceed 20 KB, their propagation time increases linearly with size. Because a protocol with a higher compression rate can compress a block into a smaller size, the protocol can improve the propagation time and decrease the fork rate. Our results suggest that protocols with lower BPT tend to have a smaller acceptable block size. Specifically, Graphene is the smallest with an acceptable size of 1.85 MB. Note that the aim of the compression protocols is to achieve a higher compression rate and a lower failure rate while preserving a high level of security and privacy protection. Protocols similar to Dino can keep the block size small (around 1 KB) no matter how many transactions are included. Their propagation time will be more stable than the other protocols whose block size increases with the number of transactions. However, further research is necessary to evaluate Dino's security and networking performance.


\section{Conclusion}

This paper evaluates six state-of-art compression protocols: Compact, XThin, Graphene, XThinner, IPFS Model and Dino, which are designed to improve Blockchain's scalability. In Experiment 1, a Monte Carlo simulation was conducted to observe the compression performance of Graphene and XThinner. Experimental results reveal that higher block space utilization by different compression protocols leads to higher transaction throughput in the Blockchain network. We limit the block size to 1 MB and use TPS as a benchmark to compare the block capacities of different compression protocols and focus on their transfer capabilities. Notwithstanding, the increasing block throughput may pose severe threats to the Blockchain under the upcoming transaction-fee regime.

Under the transaction-fee regime, miners' honest behaviour needs to be properly incentivized through transaction fees provided by the investors. While compression protocols increase the number of transactions that can be incorporated into a block, the volatility of miners' rewards can be high due to the exponential distribution of transaction arrivals and the large variation in transaction amounts. The unpredictability of miners' rewards could lead to deviant strategies that threaten Blockchain's security.

In Experiment 2, we found that the increase in scalability could lead to a significant drop in volatility when throughput was low. After reaching a critical point, better scalability leads to a gradual rise in volatility. \textcolor{black}{To better illustrate this point, we define two terms called underscaling and overscaling. Underscaling refers to the condition where the throughput is low, while overscaling condition has a high throughput. But the throughput in both conditions results in high volatility.} The experimental results also reveal that the compression protocols can result in different levels of volatility of miners' rewards when the block size is limited to 1MB. 

On the other hand, by keeping the volatility of miners' rewards within the historical range, we calculate the maximum block size for five compression solutions. Graphene has the smallest maximum block size of 1.85 MB. In contrast, Compact, XThin, and IPFS Model has a maximum block size of 4 MB. 
We hope that our analysis of Blockchain scalability and block reward volatility will pave the way for developing more advanced scalable solutions in the future.

\bibliographystyle{unsrtnat}
\bibliography{references}  






\end{document}